\documentclass[fleqn,12pt,twoside]{article}
\def\astrobj#1{#1}
\usepackage[headings]{espcrc1}
\usepackage{doublespace}

\readRCS
$Id: espcrc2.tex,v 1.2 2004/02/24 11:22:11 spepping Exp $
\usepackage{graphicx}
\usepackage[figuresright]{rotating}

\newcommand{\AmS}{{\protect\the\textfont2
  A\kern-.1667em\lower.5ex\hbox{M}\kern-.125emS}}

\title{Time evolution of simple molecules during proto-star collapse}
\author{Ankan Das,\address[CSP]{Centre For Space Physics,
        43 Chalantika,Garia Station Road, Kolkata 700084, India,
 email: ankan@csp.res.in}
       Sandip K. Chakrabarti, \addressmark[CSP]
           \address[SNBNCBS]{S.N. Bose National Center for Basic Sciences,
       JD-Block, Salt Lake, Kolkata,700098, India,
 email: sandip@csp.res.in, acharyya@csp.res.in,
}
        Kinsuk Acharyya\addressmark[SNBNCBS]{} 
       and
        Sonali Chakrabarti\addressmark[CSP] \address{Maharaja 
        Manindra Chandra College,
        20 Ramkanto Bose Street, Kolkata 700003,India,
        email: sonali@csp.res.in}}

\runtitle{Time evolution of simple molecules}
\runauthor{Das et al.}
\begin{document}
\maketitle
\begin{abstract}
\vskip 2cm
\hskip 6cm {\Large \bf Abstract}
\vskip 0.5cm
We study the formation and evolution of several molecules in a collapsing interstellar cloud 
using a reasonably large reaction network containing more then four hundred atomic and molecular 
species. We employ a time dependent, spherically symmetric, hydrodynamics code 
to follow the hydrodynamic and chemical evolution of the collapsing
cloud. The flow is assumed to be self-gravitating.
We use two models to study the hydrodynamic evolution:
in the first model, we inject matter into an initially 
low density region and in the second model, we start with a constant density 
cloud and let it collapse due to self-gravity. 
We study the evolution of the central core for both the cases.
We include the grain chemistry to compute the formation of molecular hydrogen
and carried out the effect of gas and grain chemistry at each time step.
We follow the collapse for more than $10^{14}$s (about $3$ million years) 
and present the time evolution of the globally averaged abundances
of various simple but biologically important molecules, such as  glycine, alanine etc. 
We compare our results with those obtained from observations found that
for lighter molecules the agreement is generally very good. For complex
molecules we tend to under predict the abundances. This indicates that
other pathways could be present to form these molecules or more accurate reaction
rates were needed. \\\\

Keywords: hydrodynamics; star formation; ISM; chemical evolution\\

PACS No.:  95.30.Lz; 97.10. Bt;98.38.-j; 98.62.Bj

\end{abstract}

\hspace{1.0cm}

\noindent Details of the Corresponding Author: 

\noindent Prof. Sandip K. Chakrabarti \\
\noindent S.N. Bose National Centre for Basic Sciences \\
\noindent JD Block, Salt Lake, Kolkata 700098, INDIA \\
\noindent Phone: +91 (033) 2335-5708, Fax : +91 (033) 2335-3477\\
\noindent e-mail: chakraba@bose.res.in, Mobile: +919903120700\\

\newpage

\section{Introduction}

More than $125$ species of molecules have been observed in the interstellar 
clouds and star forming regions. Among them, over half are organic.
Serious efforts have been made over the years to investigate
the formation of such molecules in cool interstellar clouds in frigid 
conditions (Hasegawa et al., 1992; Hasegawa and Herbst, 1993; Leung et al., 1984;
Prasad and Huntress, 1980a, 1980b). 
It is now
quite certain that the most important building block, namely, the molecular hydrogen 
($H_2$) and some of the other lighter molecules must be produced in the presence 
of grains (Gould and Salpeter, 1963; Hollenbach and Salpeter, 1971; 
Hollenbach et al., 1971).
Several analytical and numerical 
works have successfully shown how the molecular hydrogen may have been
produced (Biham et al., 2001).
A number of results are present in the 
literature where hydrodynamic and chemical evolutions have been attempted 
simultaneously. For example, Shalabiea and Greenberg (1995)
used the pseudo as well
as partially real time-dependent models for the hydrodynamical
evolution. In the pseudo time-dependent method, they assumed a constant 
density and temperature of the cloud using which the chemical 
evolution was computed. In their time-dependent model, 
they included the density and temperature
variations throughout the cloud. However, in their initial approach to the
time-dependent modelling, they assumed a constant temperature
but allowed only the density to vary. 
Ceccarelli et al. (1996) used the ``inside-out",
isothermal, spherical collapse model of 
Shu (1977) and coupled it with a time-dependent chemical evolution code. 
They included the heating and the cooling processes with an
emphasis on the line emission. 
Shematovich et al. (1997) used Zeus 2D code which included the heating and the cooling. 
They present 1D 
hydrodynamic and chemo-dynamical evolution of the proto-stellar cloud 
illuminated by the diffused interstellar UV radiation. They solved the 
equations of chemical kinetics, hydrodynamics and thermal balance 
simultaneously. In 
Lim et al. (1999) 2D numerical code was developed using the adaptive 
grid technique. Here, $454$ reactions among $42$ atomic and molecular 
chemical species were taken including the basic elements like $H$, $He$, $C$, $N$, $O$ 
and a representative low ionization potential metal $Na$.
At each grid point, the chemical evolution was followed by a calculation of the 
reaction rates using the local conditions obtained
from the hydrodynamical flow. They primarily concentrated on the diffused 
clouds and emphasized the interfaces of the interstellar media and 
resulting dynamical mixing. 
Aikawa et al. (2005) studied time-dependent evolution of 
Bonner-Ebert spheres by assuming clouds having a specific parameter 
$\alpha$ which is the ratio of the gravitational force to the pressure 
force. Recently, Acharyya et al. (2005)
solved the Master equations and rate equations of 
Biham et al. (2001) for various cloud parameters and followed the 
evolution of $H_2$ as a function 
of time. Both this work and the earlier works of 
Chakrabarti and Chakrabarti (2000a) employed steady state matter distribution and 
assumed that the density and the temperature distributions at a given radial 
distance do not change with time. 
Chakrabarti and Chakrabarti (2000a) used a 
large number of species and 
the reaction rates were taken from the UMIST data base. Some of the reaction 
rates which were not available in the literature were assumed to be similar 
to other two body reactions. Subsequently, these new and assumed reaction rates
were parametrized (with reaction rates up to a thousand times
smaller compared to Chakrabarti and Chakrabarti (2000a)
to include the effect of the size of the reactant
molecules (Chakrabarti and Chakrabarti, 2000b).
It was shown that even under frigid and tenuous conditions 
of the interstellar media, a significant and perhaps a detectable amount of
simple amino acids and even important ingredients of DNA molecule (such as
adenine) may form. 
Ceccarelli et al. (2000) estimated the upper 
limit of the abundance of glycine to be about $10^{-10}$ (cooler outer cloud) to 
$7\times 10^{-9}$ (hot core). 
Kuan et al. (2003) estimated the fractional 
abundance of glycine to be $2.1 \times 10^{-10}$ for Sgr B2, $1.5 \times 10^{-9}$ for 
Orion, and $2.1 \times 10^{-10}$ for W51. 
These numbers are comparable to what was predicted 
in Chakrabarti and Chakrabarti (2000a),
however, there are clearly some debate on the possible pathways for the formation of 
glycine with the route followed in the Chakrabarti and Chakrabarti (2000a, 2000b).
Similarly, there are also some debate on whether glycine is actually 
observed (Hollis et al., 2003; Snyder et al., 2005).
A few other relevant results which may 
be mentioned in passing are as follows. 
Tarafdar et al. (1985) presented a model of
chemical and dynamical evolution of isolated, initially diffused and 
quiescent interstellar clouds. A semi-empirically derived dependence
of the observed cloud temperatures on the visual extinction and density was 
used in this work. 
Sorrell (2001) outlined a theoretical model for the 
formation of the interstellar amino acids and sugars. In this model, 
first ultraviolet photolysis creates a high concentration of free radicals 
in the mantles and the heat input due to the grain-grain collision causes 
radicals to react chemically with another to build complex organic molecules.
Bernstein et al. (2002) reported a laboratory demonstration that the complex 
bio-molecules like glycine, alanine etc. are naturally formed from the 
ultraviolet photolysis of the interstellar grains. 
Munoz Caro et al. (2002) report the 
detection of amino acids even at room temperature on an interstellar ice analogue 
that was irradiated with ultraviolet light in a high vacuum at 12K.
Altogether sixteen amino acids were identified. The results demonstrate 
that a spontaneous generation of amino acids in the interstellar medium 
is possible, supporting the suggestion that prebiotic molecules could have 
been delivered to the early earth by cometary dust, meteorites or 
interplanetary dust particles.

In this backdrop of this observational and experimental status, we carry out
our investigation by improving earlier work by first incorporating the accurate grain 
chemistry as elaborated in 
Acharyya et al. (2005) and then by actually combining 
the results of a time dependent hydrodynamics code with the chemical evolution code 
to see how the abundances vary with the grid locations. We also chose initial 
conditions very much different from the earlier studies. We used two different 
but realistic models. In one model (Model A), the cloud matter is injected through the
grid boundary at a constant speed and the cloud as well as the core are 
allowed to form {\it ab initio}. In the other model (Model B), the computational 
grid area is chosen to be the central part of a much larger spherical cloud of 
constant density and temperature. The rate at which the interstellar matter enters 
into the grid (i.e., the rate at which the larger cloud is evacuated) depends 
on the gravitational pull between the inner cloud within computatioal grid 
and the outer cloud outside of the computational grid. In following the
chemical evolution, we used the available  standard reaction rates for most of the
reactions, but the rates of some of the very complex molecule formation 
are still very much uncertain and as such we present the evolution of the 
mass fractions only for simpler bio-molecules. In the next section,
we present the hydrodynamic equations which govern the cloud collapse. In \S 3, 
we present the hydrodynamical and chemical evolution. In \S 4, we 
discuss in detail the nature of the cloud models which we simulate and how 
the chemical evolution code is used in conjunction with the results of the 
hydrodynamic simulation. In \S 5, we present the results.
Finally, in \S 6, we make concluding remarks.

\section{Hydrodynamic Equations Governing the Cloud Collapse}

The time dependent simulations have been carried out by various authors 
using both the Lagrangian and Eulerian methods. For instance, 
Prasad et al. (1991) and Tarafdar et al. (1985) used a Lagrangian scheme with a 
semi-empirical approach for the energy equation.
Shalabiea and Greenberg (1995) 
used a partially time dependent model in which only the densities are 
allowed to be time dependent during collapse of the cloud. We have used
a finite difference Eulerian scheme (upwind scheme) in which we difference the
one dimensional Euler equations along the radial grid. We consider a self-gravitating, 
collapsing, spherically symmetric gas cloud. We choose the co-ordinate 
system to be ($r, \theta, \phi$) with origin at the center of the proto-star. 
Since we would be interested in the spherical case, we ignore the $\theta$ 
and $\phi$ component motions and concentrate only on the radial equation in 
this paper. The Eulerian equations of hydrodynamics written in spherical 
coordinates are given by:

$$
\frac{\partial \rho}{\partial t} + \frac{1}{r^2}{\frac{\partial 
(\rho v_r r^2)}{\partial r}} = 0,
\eqno (1)
$$
and
$$
\frac{\partial \rho v_r}{\partial t} + \frac{1}{r^2}{\frac{\partial
 (\rho v_r^2 r^2)}{\partial r}} = - (\rho \frac{\partial \Phi}
{\partial r} + \frac{\partial p}{\partial r}).
\eqno (2)
$$
Here, $\Phi$ is the gravitational potential [$=-GM(r)/r$] where, $M(r)$ is 
the mass of the cloud inside radial distance $r$.

We assume the ideal gas equation to be,
$$
p=\rho k T/\mu m_p,
\eqno (3)
$$
where, $k$ is the Boltzmann constant, $T$ is the local temperature, $\mu$
is the mean molecular weight, $m_p$ is the proton mass.

\section{Hydrodynamical and Chemical Evolution}

The equations presented in the previous Section have been solved  to study 
full time-dependent  behaviour of the spherical flow. The actual solution 
depends on the model flow. In the present paper we used two types of initial 
conditions for the evolution of two hydrodynamic models. These are coupled 
with chemical evolution code to obtain time dependent chemical composition.

\subsection{Hydrodynamical Models}

(a) Model A: Here we start with a grid of size $r_{out}$. We assume that 
initially the cloud contains a negligible amount of mass. At the outer 
boundary, we inject matter at a constant rate of $\frac{d M}{d t} = 4\pi \rho_{out} 
v_{out} r_{out}^{2}$, where, $\rho_{out}$ is the injection density and $v_{out}$ 
is the injection velocity at the outer boundary ($r=r_{out}$). Using this model, 
we mimic the formation of the cloud itself from a supply of matter from a large 
reservoir of diffused gas.

The inner boundary is chosen at $r=r_{in}$. The distance $r_{out}-r_{in}$
is divided into $N=100$ logarithmically equal spaced grid points.
At each time step, the mass of the core is dynamically updated by the 
amount of matter that is getting inside $r_{in}$. The rate of increase 
of $M_{core}$ is,
$$
\frac{d M_{core}}{d t}=4\pi \rho_{in} {v_r}_{in} r_{in}^2 ,
\eqno{(4)}
$$
where, $\rho_{in}$ is the density at the inner grid point, ${v_r}_{in}$ is 
the velocity at the inner grid point, $r_{in}$ is the radial distance of the 
inner grid point from the center of the molecular cloud. Similarly, the 
mass of the cloud is also dynamically updated. The flow
is assumed to be isothermal throughout and $T=10$K is assumed for concreteness.

(b) Model B: In this model, we assume that we have a finite sized $r=r_{res}$
reservoir of matter and we are considering only the inner region of size $r_{out}$ for 
computational purpose. The matter is coming from the reservoir and it is 
evacuated to form the star itself. The inward velocity is computed 
self-consistently from the gravitational pull between the matter 
inside $r_{out}$ and that outside of $r_{out}$. The whole cloud is assumed 
to be isothermal (with $T=10$K) and has a constant density 
$\rho_{int}$ throughout. The size of the whole cloud is $r_{res} >> r_{out}$. 
We assume that $M_{total}=M_{core}+M_{cloud}+M_{out}$, where, $M_{core}$ 
is the core mass within $r=r_{in}$ which will increase at the same rate as in eq. (4), 
$M_{cloud}$ is the mass of the cloud within the computational grid,
i.e., in between $r_{in}$ and $r_{out}$ and $M_{out}$ is the mass in between the radius 
$r_{res}$ and $r_{out}$. Thus, $M_{out}$ is the reservoir mass which is
depleted as matter is injected within $r=r_{out}$. The injection velocity 
is dynamically calculated from the following way. First, we note that an 
approximate expression for the acceleration of matter at the outer 
edge of the computational grid could be taken as, 
$$
f_{out} \sim \frac{G(M_{cloud}+M_{core})}{{r_{m}}^2},
\eqno{(5a)}
$$
where, $r_m= (r_{res} +r_{out})/2$. Hence the injection velocity is,
$$
v_{out} \sim \sqrt{f_{out} r_{m}},
\eqno{(5b)}
$$
and the average density of the injected matter is,
$$
\rho_{out} = \frac{M_{out}}{V_{out}},
\eqno{(5c)}
$$
where, $V_{out}=4/3\pi(r_{res}^3 - r_{out}^3)$. By this process, the 
large cloud would be gradually evacuated as the star is formed. 
The cloud mass $M_{cloud}$ inside $r_{in}<r<r_{out}$ will increase at the rate,
$$
\frac{d{M_{cloud}}}{dt}= 4\pi(\rho_{out}v_{out}r_{out}^2-\rho_{in}v_{in}r_{in}^2)
\eqno{(6)}
$$
where, $v_{in}$ and $\rho_{in}$ are respectively the velocity and density 
at the inner boundary $r_{in}$.  

\subsection{Chemical Evolution}

The chemical evolution code is similar to that used in 
Chakrabarti and Chakrabarti (2000a, 2000b)  
except in one 
very important way, namely, the way we handle the grain chemistry. 
In Chakrabarti and Chakrabarti (2000a, 2000b) a
constant rate of recombination of $H+H \rightarrow H_2$ was chosen 
following the prescription of Miller et al. (1997).
In our present work, however, we incorporate the grain chemistry properly for the
production of $H_2$ at each radius. Here, we followed the 
techniques used in Acharyya et al. (2005).
We used only the olivine grains for simplicity. We also incorporate the 
grain size distribution according to Weingartner and Draine (2001a, 2001b)
and sub-divided the grains into three major types 
(a) $5 $ $\AA$, (b) $75$ $\AA$, and (c) $0.2\mu$ so that the calculation 
of $H_2$ becomes faster. Whether or not the Master equation or the rate 
equation would be used depends on the flux of the infalling hydrogen
(Acharyya et al., 2005) on the grain surface. We take the reaction rates from 
Miller et al. (1997) which supplies the rate constant $k$ for a two body reaction as:
$$
k=\alpha {(T/300)^\beta}exp(-\gamma /T)  {\rm cm}^3 {\rm s}^{-1} ,
\eqno (7)
$$
where, $\alpha$, $\beta$, $\gamma$ are constants.
The reactions for which the rates were not available in the UMIST data base,
we take a conservative value of $\alpha =10^{-10}$ cm$^{-3}$ s$^{-1}$,
$\beta = \gamma = 0$ like any other typical two body (open shell) reactions
even though we may be using neutral-neutral reactions. The chemistry of
bio-molecule formation in space is very much unknown and it is difficult
to quantify the rate with any certainty. Thus the results obtained using 
these rates may have some errors. 

Two models have been run
with different initial abundances of carbon. Generally, we keep the initial 
composition as was kept in Miller et al. (1997)
(after converting to mass fractions as 
in Chakrabarti and Chakrabarti, 2000a), 
i.e., H:He:C:N:O:Na:Mg:Si:P:S:Cl:Fe = $0.64$:$0.35897$:$5.6\times
10^{-4}$:$1.9\times 10^{-4}$:$1.81\times 10^{-3}$:$2.96\times 10^{-8}$:
$4.63\times 10^{-8}$:$5.4 \times 10^{-8}$:$5.79\times10^{-8}$:$4.12\times
9^{-7}$:$9\times 10^{-8}$:$1.08\times10^{-8}$. This Set is our `Set 1' initial
abundance. In `Set 2' initial abundance,
we used a higher amount of carbon to simulate the chemical evolution
inside a collapsing cloud in the environment of evolved stars.

While we include all the reactions given in the UMIST data base, we
wish to remind that since the interstellar medium is cooler, 
the exothermic reactions should be more favourable.
The following major types of the reactions are included 
(Tielens, 2005):

\noindent {\bf (i)  Photochemistry:}

The UV photons present in the diffuse ISM are a dominant destruction
agent for small molecules. Inside a cloud, the radiation field will be
attenuated by the dust. In the unshielded radiation field this type of
reactions have a typical rate of $10^{-9}$  s$^{-1}$. These
reactions mainly occur in the outer edge of the molecular cloud 
where UV photons are profuse in number. Deep inside the cloud some 
UV photons can be created due to radiative association, which could
also lead to this type of reactions. The rate is calculated by,
$$
\kappa_{pd}= a exp[-b A_v],
$$
where, $A_v$ is the visual extinction due to the dust, $a$ is the unshielded 
rate and $b$ is the self-shielded rate. An example of this kind of reaction is,
$$
CH + h\nu \rightarrow C + H .
$$

\noindent {\bf (ii) Neutral-neutral:}
This type of reactions often possesses appreciable activation barriers 
because of the necessary bond breaking associated with the molecular 
re-arrangement. This type of reactions is important when the gas is warm,
e.g., in stellar ejecta, in hot cores associated with proto-stars, in dense 
photo-dissociation regions associated with luminous stars, or in the 
post-shock regions. The rates are generally around 
$4 \times 10^{-11} \ cm^3 \ s^{-1}$ when no activation barrier is present. 
Reaction rates are calculated as,
$$
\kappa=\alpha(T/300)^{\beta}exp(-\gamma/KT),
$$
where, $\alpha,\  \beta,\ \gamma$ are used to calculate the rate coefficients.
One example of this type of reaction is,
$$
H_2 +O \rightarrow OH +H .
$$
In general, the only neutral-neutral reactions that occur in the
cold conditions of the dark clouds are those involving atoms
or radicals, often with non-singlet electronic ground states that
do not have activation barriers. One example of this type of reaction is,
$$
C+OH \rightarrow CO + H .
$$

\noindent {\bf (iii) Ion-molecule reactions:}
Exothermic ion-molecule reactions occur rapidly because the strong
polarization-induced interaction potential can
be used to overcome any activation barrier energy involved.
In the exothermic direction, this kind of reactions have a typical rate of
$2 \times 10^{-9} \ cm^3 \ s^{-1}$ and the reaction rates are in the form
$$
\kappa=\alpha.
$$
One example of this kind of reaction is,
$$
H_2^+ + H_2 \rightarrow H_3^+ + H .
$$

\noindent {\bf (iv) Charge transfer reactions:}
This type of reactions is of great importance for setting the ionization
balance in the HII region. The charge exchange between O and $H^+$ is a 
very important reaction in the ISM because it ionizes the
oxygen which is then able to participate in the chemistry of the ISM.
This type of reactions has a typical rate of $10^{-9} \ cm^3  \  s^{-1}$ in the
exothermic direction. An example is:
$$
H^+ + O \rightarrow H + O^+ .
$$

\noindent {\bf (v) Radiative association reactions:}
In this type of reaction, the product after the collision of two species
is stabilized through the emission of a photon. The reactions of this type have
highly reaction specific rates. The reaction rates are given by,
$$
\kappa=\alpha .
$$
One example of this type of reaction is,
$$
H+C \rightarrow CH + h\nu .
$$

\noindent {\bf (vi) Dissociative electron recombination reactions:}
This type of reactions involves in the capture of an electron
by an ion to form a neutral in an excited electronic state
that can dissociate. Typical rate of such a reaction is
around $10^{-7} \  cm^3 \  s^{-1}$ and the rate coefficient is calculated by,
$$
\kappa=\alpha(T/300)^{\beta}.
$$
One example of this kind of reaction is,
$$
OH^+ + e \rightarrow  O + H .
$$

\noindent {\bf  (vii) Associative detachment reactions:}
In this case, an anion and an atom collide and the neutral product stabilizes
through an electron emission. This type of reactions has a rate of around
$10^{-9} \  cm^3 \  s^{-1}$ in the exothermic direction.
$$
H+H^- \rightarrow H_2+e .
$$

\noindent {\bf (viii) Collisional association:}
In laboratory settings, three-body reactions generally dominates chemistry,
$$
A+B+M \rightarrow AB+M
$$
with rates $ \approx 10^{-32} cm^6 \ s^{-1}$. These reactions generally 
have very little importance in the astrophysical environment except for dense
gas near stellar photosphere or in dense($\approx 10^{11} cm^{-3}$) 
circumstellar disks. 

We do not give the reaction pathways since there would be too many of them. 
Generally, pathways given in Chakrabarti and Chakrabarti (2000a, 2000b)
have been used for synthesis of those complex molecules which are 
absent in the UMIST data base.
                                
\section{Results}
Armed with a working time-dependent hydrodynamical evolution code and a 
chemical evolution code, we ran two models (Models A and B) of the 
interstellar cloud collapse. In both the Models we choose, 
$r_{in}=3.98 \times 10^{13}$cm and use $100$ grid points
which are spaced equally in the logarithmic scale along
the radial direction. We assume that 
anything going inside $r_{in}$ will increase the mass of core while the
the radius of the core remains $r_{in}$. In Model A, we chose 
$r_{out}=3.54 \times 10^{18}$cm and we ran for both the Sets 1-2 as the initial 
compositions. We started our simulation with $\rho_{int}
=10^{-27}$ gm cm$^{-3}$ in all the grids (except on the boundary, see below).
Thus, we start with a core of a negligible mass of
$(4/3)\pi r_{in}^3\rho_{int}= 2.64 \times 10^{14}$gm.
We inject matter at $v_{int}=50$cm s$^{-1}$ at each 
grid point in the outer boundary.
                                                                                                              
In Model B1, we combine the hydrodynamical code and the chemical evolution code
using only the Set 1 as the initial composition. Here, we choose 
$r_{res}=10^{20} cm$, $r_{out}=3.16 \times 10^{18} cm$ and 
$r_{in}=3.98\times 10^{13}$ cm and use equal logarithmic radial
spacings as before. At the start of the simulation $\rho_{int}=10^{-22}$ 
gm cm$^{-3}$ and $v_{int}=10$ cm s$^{-1}$ at each grid point. Initially, 
the core mass was $(4/3)\pi r_{in}^3\rho_{int} = 2.64 \times 10^{19}$ gm 
in this Model. In Model B2, we choose $r_{res}=3.16 \times 10^{18} cm$,
$r_{out}=3.16 \times 10^{17} cm$ and other conditions are same as above.

As far as the chemical evolution is concerned, we use a large network 
of about $4000$ 
reactions and solve $422$ equations simultaneously each of which is to 
update the mass fraction of one single specie. However, in order to 
simplify the chemical evolution and to save the computational time, we do not run 
the chemical evolution code at each grid at each time step. Instead, we divide 
the entire region under consideration in ten logarithmically equal spaced 
zones along the radial direction. Appropriately weighted averaged density 
is used in each such zone. The initial abundance of matter in each zone is 
updated using the existing abundance in that zone plus the abundance of the 
upstream matter advecting in and minus the abundance of the downstream matter
advecting out. The time step in chemical evolution is totally dictated by the
fastest reaction in the network. The evolution process is continued till 
the end of the hydrodynamic simulation. While we have time evolution of the 
abundance of each species at each zone, we plot only the global average of the 
abundances just to give an idea of how the average abundance is evolving.

\subsection {Results of Models A and B with Set 1 as initial composition}

In this Model, we choose $\rho_{out}=10^{-22}$\,gm cm$^3$, 
$v_{out}=10^4$\,cm s$^{-1}$. Thus, the rate of injected matter is given by 
$\dot M = 4\pi \rho_{out}v_{out} r_{out}^2= 1.57 \times 10^{20}$\,gm s$^{-1}$. 
This matter moves in due to self-gravity of the cloud and the attraction of 
the core. The simulation was carried out till a few times $10^6$\,yr. Figure 
1a shows how the velocity profile of the flow changes with time. Time 
(in years) is marked on each curve. A flow which began with a constant 
velocity, eventually assumes an almost steady state ($v(r) \sim r^{-\alpha}$ 
with $\alpha \sim  4/5$). The evolution of the density distribution is shown
in Figure 1b. Here too, an initially constant density distribution assumes a 
power-law distribution of $\rho(r) \sim r^{-3/2}$ toward the end of the 
simulation. Time (in years) is marked on each curve. Note that once the 
core becomes massive, it starts to evacuate the grids and thus the density over the 
entire cloud gradually decreases as is shown by the dot-dashed curve drawn 
at $\sim 6.33 \times 10^6$\,yr. After about $t=10^{13}$\,s,
i.e., about $3\times 10^5$\,yr, densities on each grid started increasing 
after the empty grids are filled in by the inflowing matter. This is shown 
in Figure 1c where the time evolutions of the densities on the 50th grid 
(solid curve) and the innermost grid (dotted curve) are shown. Towards the 
end, we note that the densities at these grids gradually decrease as the cloud 
starts to become empty. Meanwhile, the core starts to grow because of the
mass accretion during this period. Figure 1d shows how the mass of the core 
is evolving.

\begin {figure}
\centering
\begin{tabular}{ccc}
\includegraphics[width=.4\linewidth]{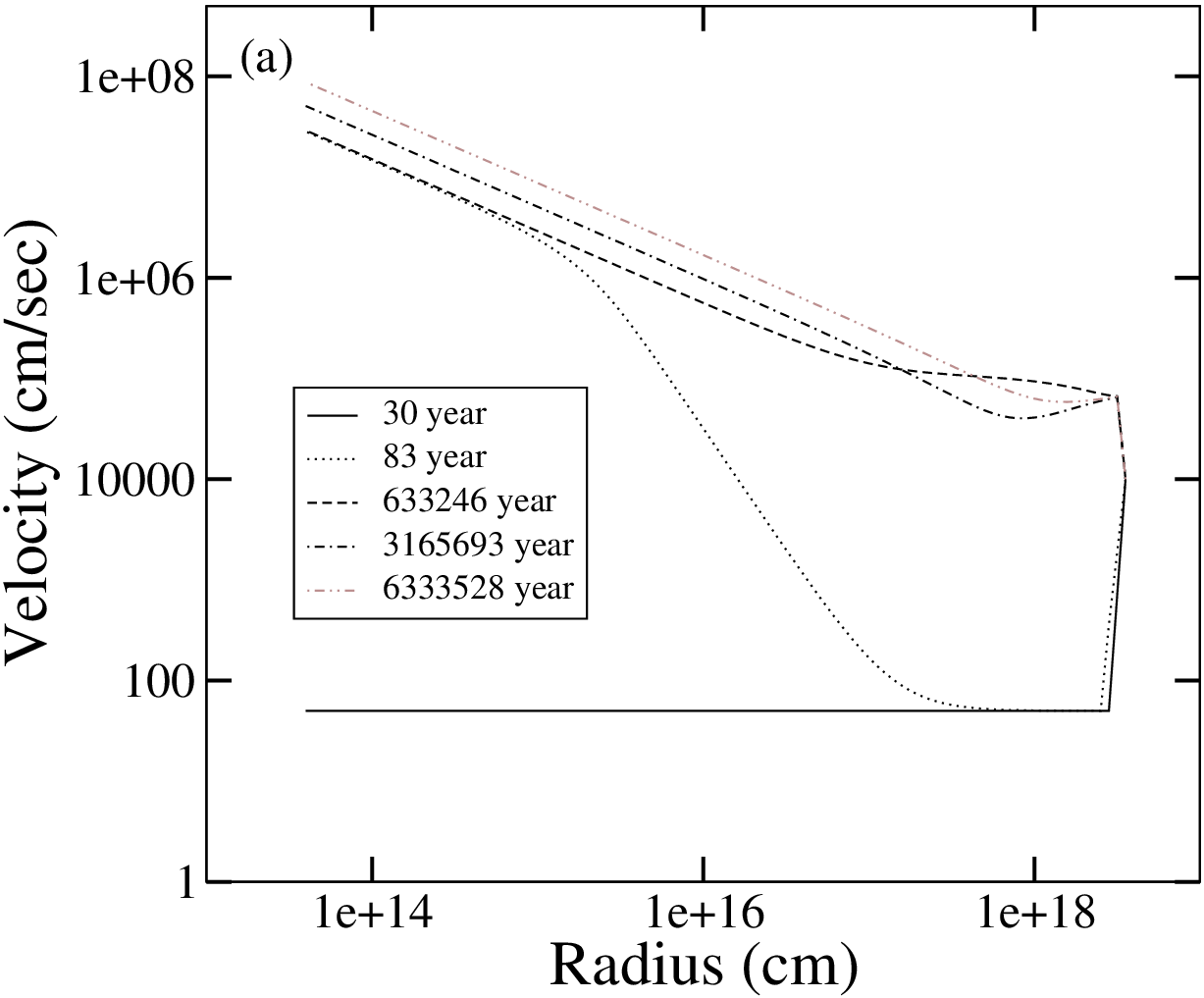} & \hspace{1cm} &
\includegraphics[width=.4\linewidth]{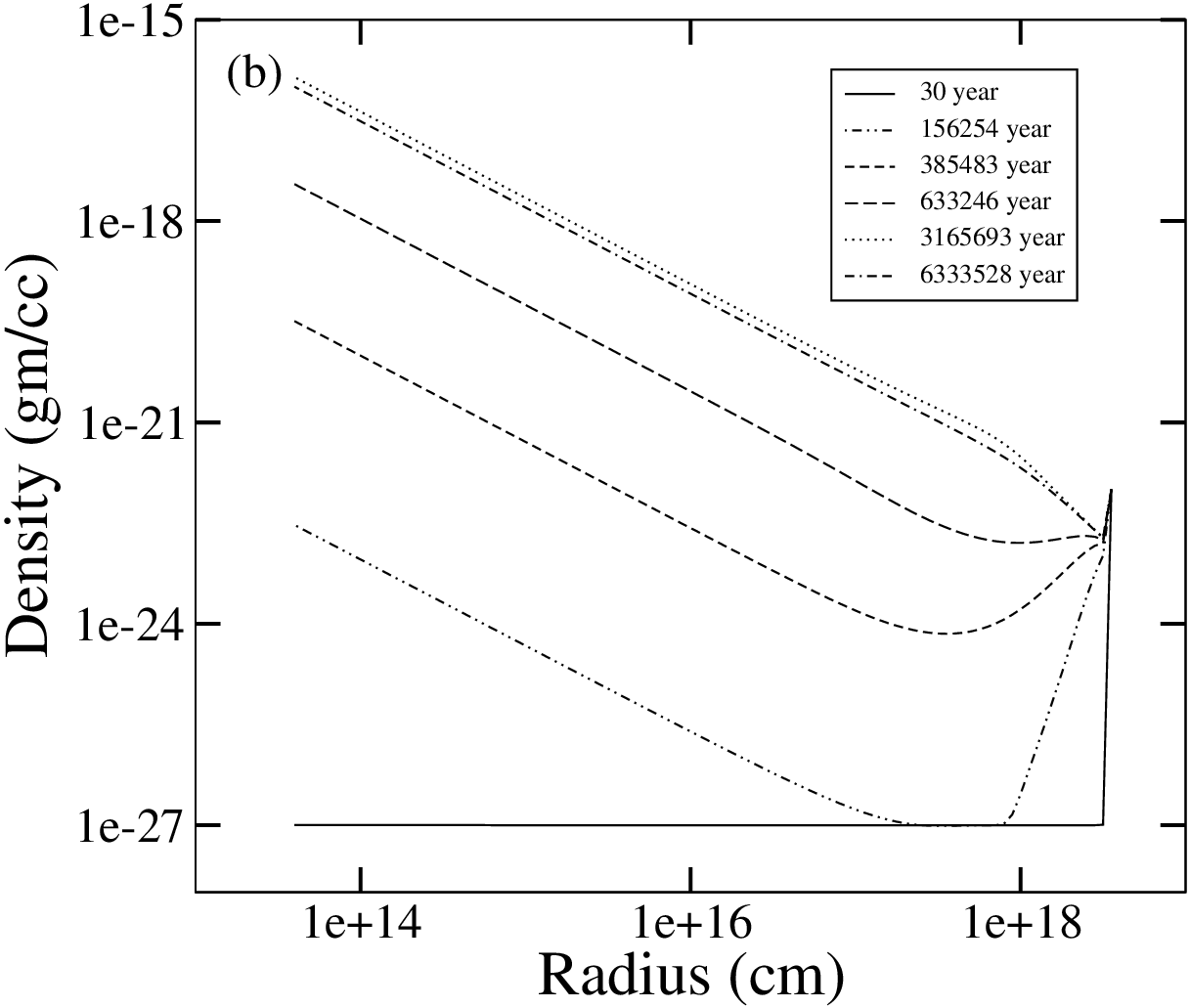} \\[1cm]
\includegraphics[width=.4\linewidth]{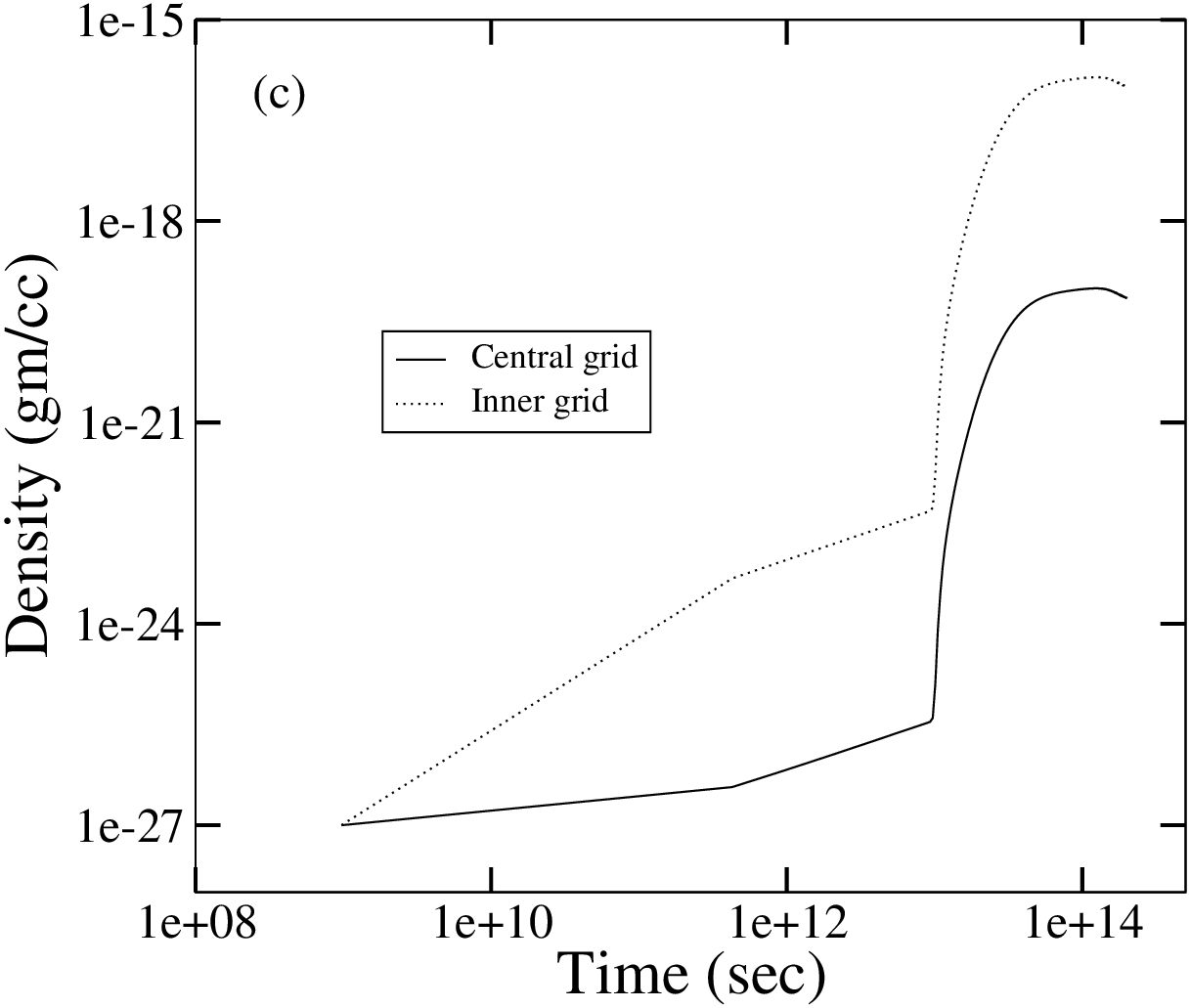} &&
\includegraphics[width=.4\linewidth]{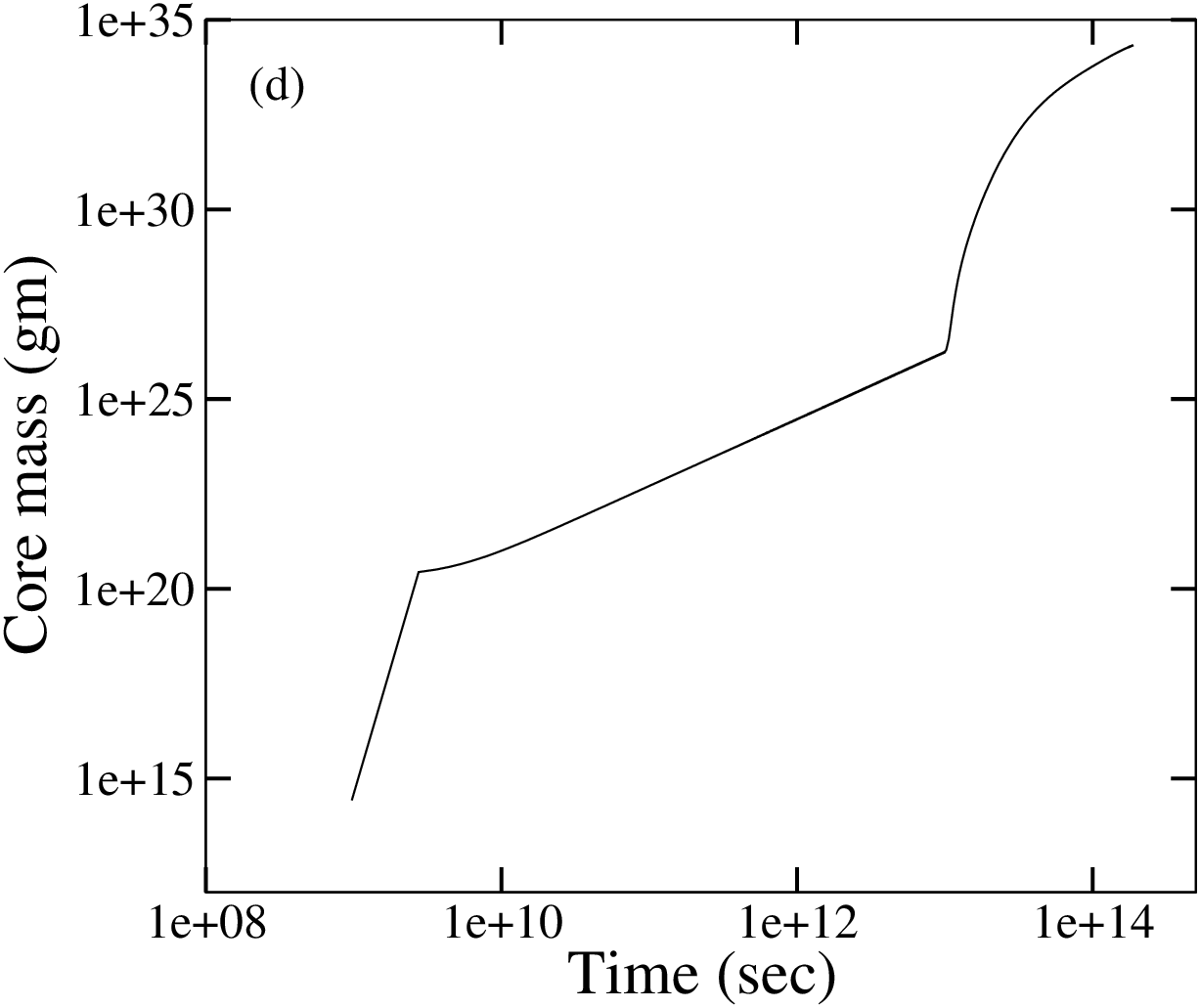}
\end{tabular}
\caption{Evolutions of (a) the velocity and (b) the density 
distribution inside the collapsing cloud of a Model A simulation.
Also shown are the evolutions of the density at (c) the middle (solid) grid
and the innermost grid (dotted) and (d) the growth of the proto-stellar core 
with time. The densities in the grid have a sharp rise after the initially 
empty grid is filled in by the infalling matter at about $10^{13}$s. From 
this time onward, the core also starts growing rapidly.}
\end{figure}

\begin {figure}
\centering
\begin{tabular}{ccc}
\includegraphics[width=.4\linewidth]{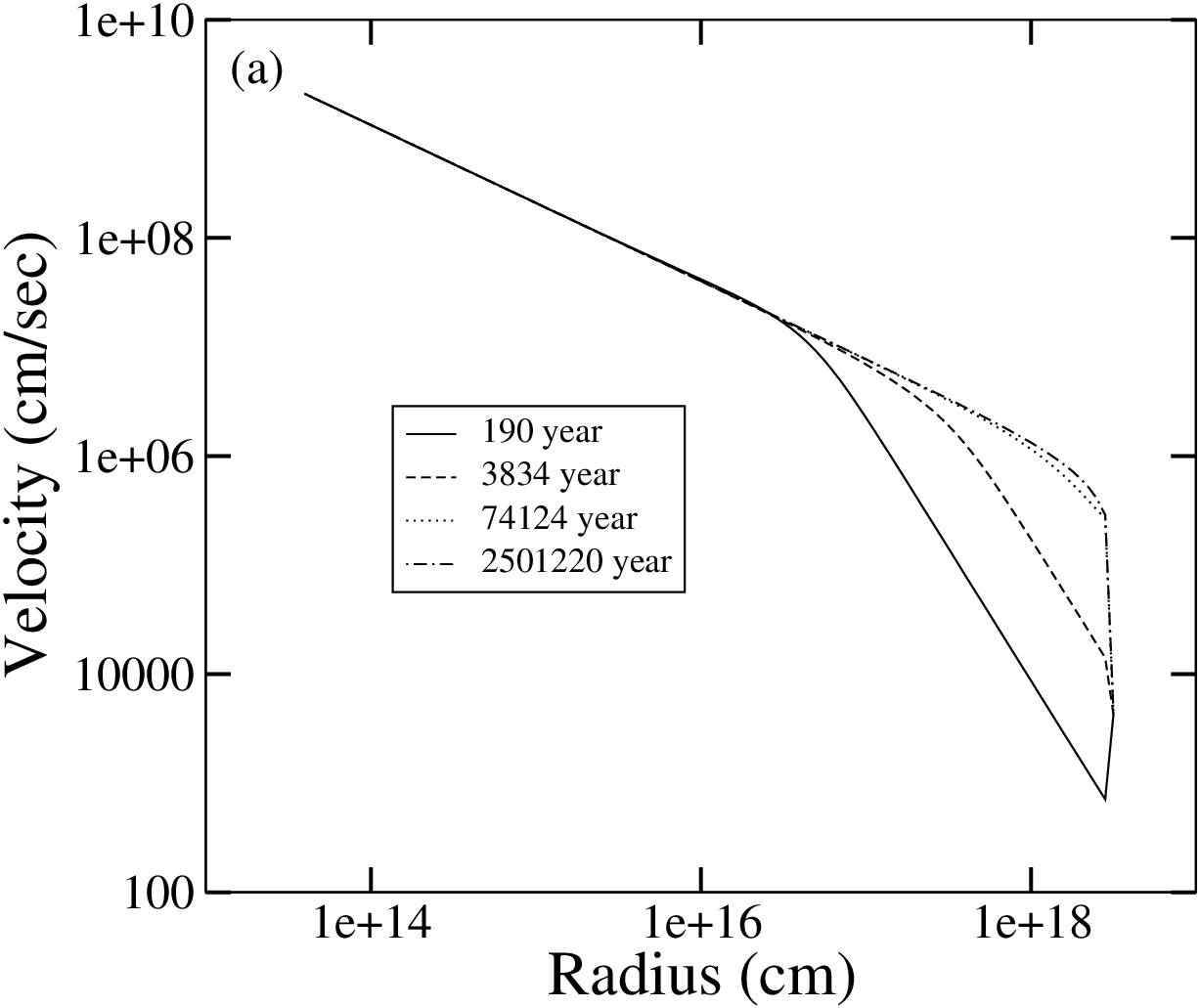} & \hspace{1cm} &
\includegraphics[width=.4\linewidth]{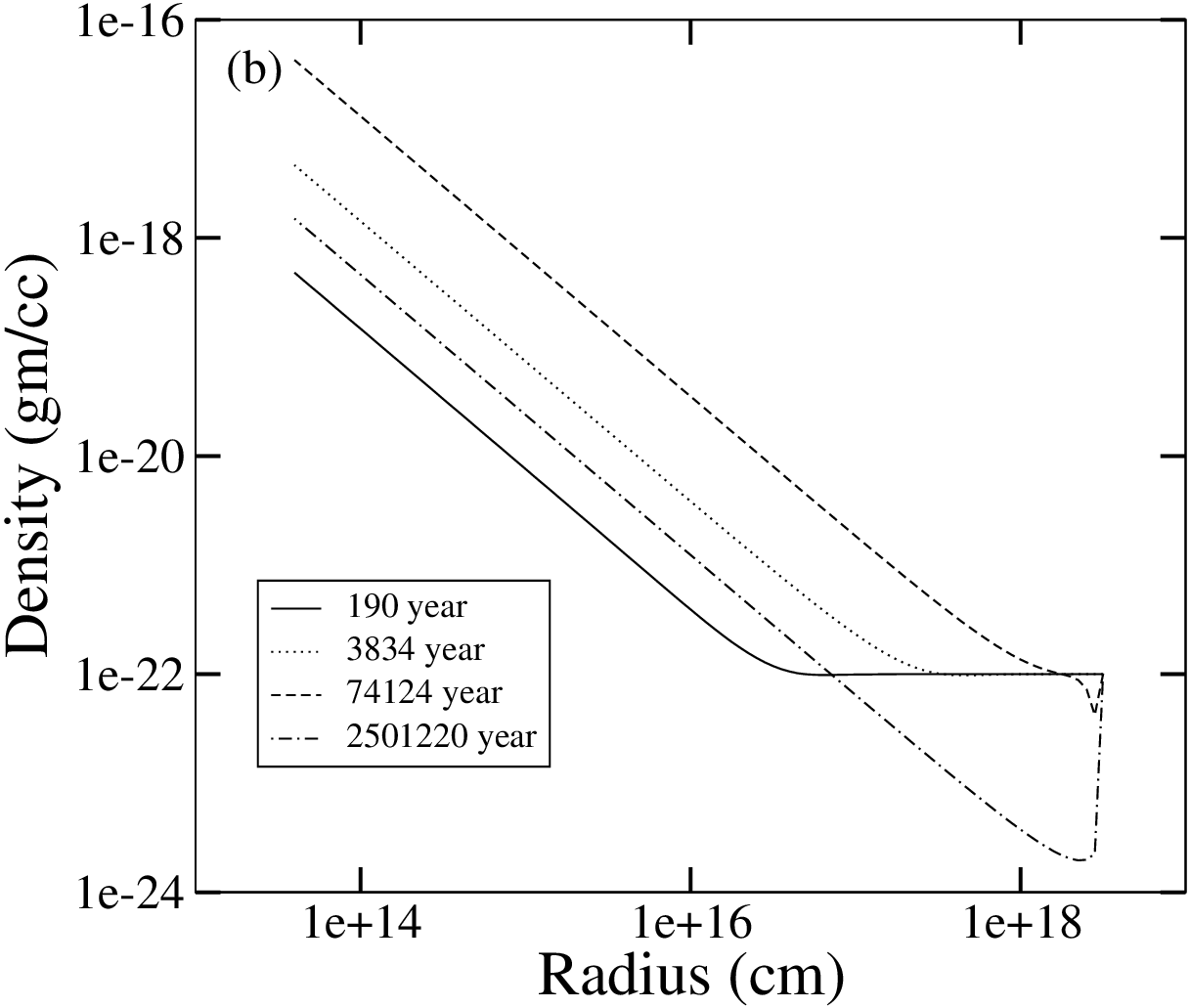} \\[1cm]
\includegraphics[width=.4\linewidth]{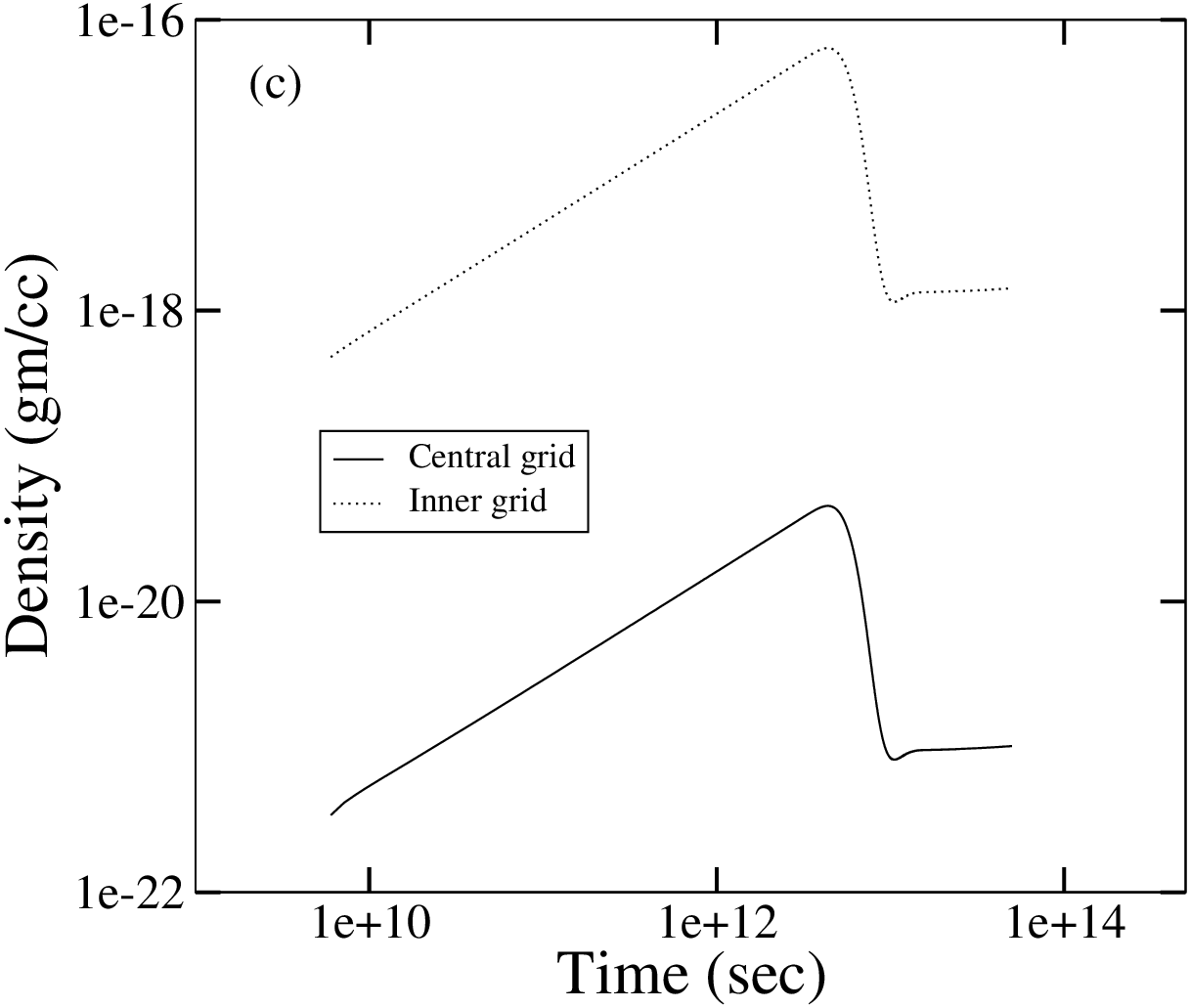} &&
\includegraphics[width=.4\linewidth]{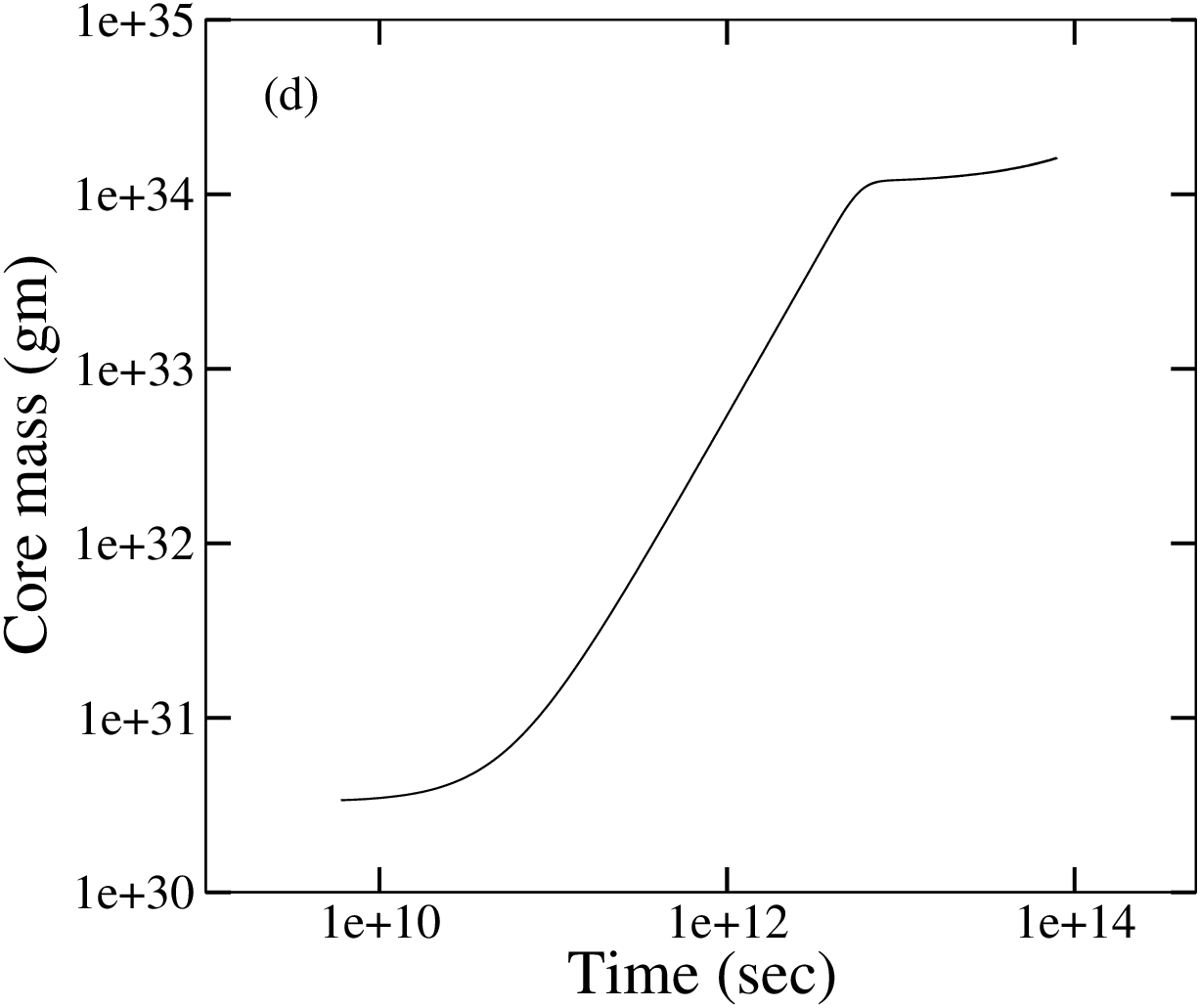}
\end{tabular}
\caption{Same as Figs.1(a-d) except that Model B1
is used for simulation. In this case, the depletion
of matter from a cloud of fixed mass takes place and the core mass of the 
cloud goes up initially rapidly and slowly afterwards.}
\end{figure}

In Model B1, the velocity assumes almost steady state 
after some initial transient time as before.
The evolution of velocity is shown in Figure 2a. In Figure 2b, the
evolution of density is shown. Initially, the density of the cloud 
increases and new matter is injected from the cloud external to the grid, 
but as the time proceeds, the core grows
while the external cloud is evacuated. This results in eventual decrease 
in density of the cloud as shown by the dot-dashed curve in Figure 2b. 
Similarly, from Figure 2c, it is clear that the density at the middle grid 
(solid curve) and the innermost grid (dotted curve) will also go down due to 
the growth of the core mass. In Figure 2d, the evolution of core is shown. Unlike 
in the previous model, the core mass starts growing from the very beginning of 
the simulation, but the rate of growth is slowed down when the cloud mass is 
depleted.

Our Model A and Model B are two complimentary models. In Model A, the initial 
grid was almost empty and thus the core mass was growing slowly. Eventually the
matter injected at the our boundary caught up and the core grew rapidly. 
In Model B, the initial grid was full. After it is evacuated, matter from the 
reservoir is accreted slowly. Thus, initially, the core grows rapidly but the 
rate of growth slowed down in the late phase.

\begin {figure}
\centering
\begin{tabular}{ccc}
\includegraphics[width=.4\linewidth]{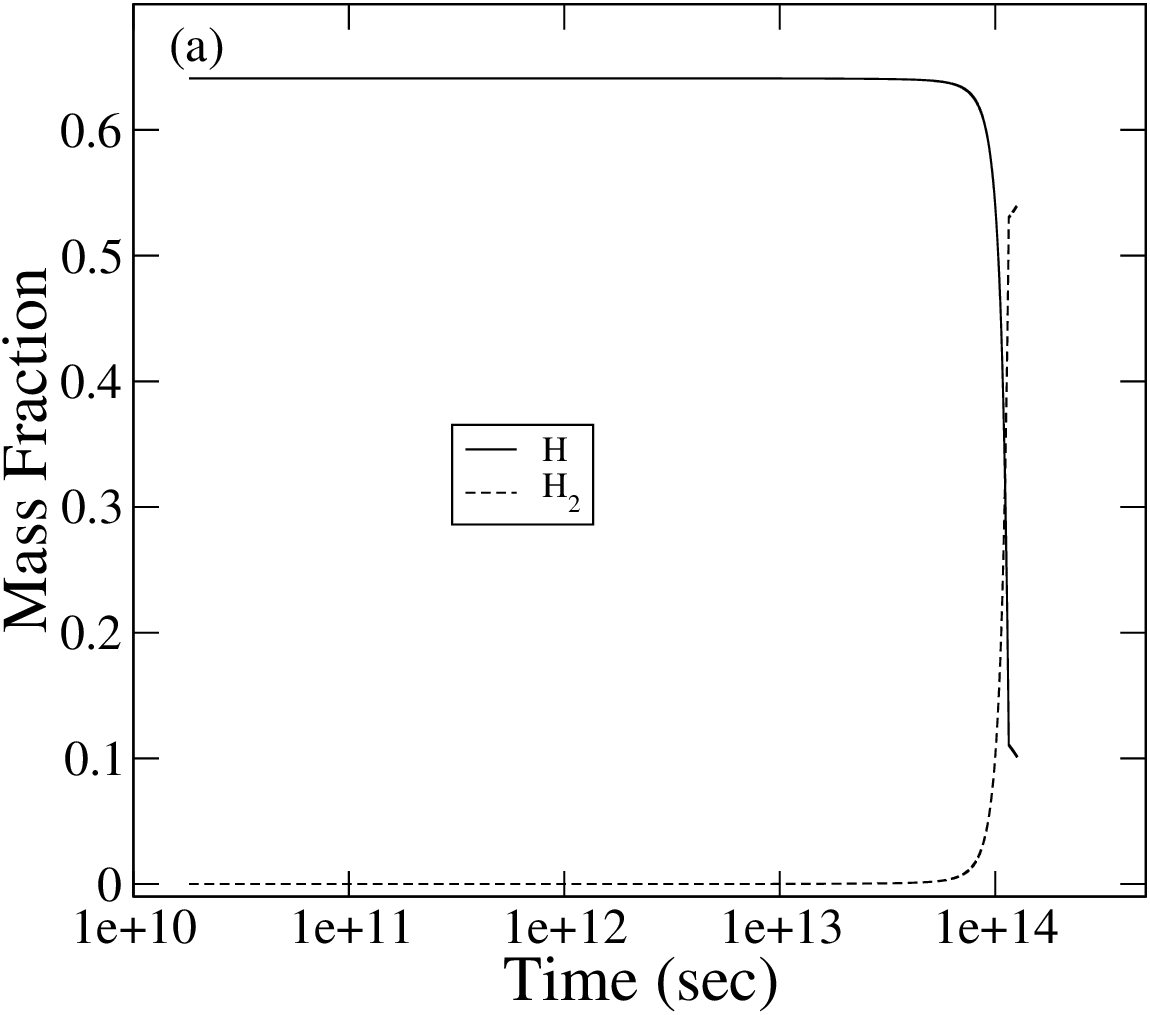} & \hspace{1cm} &
\includegraphics[width=.4\linewidth]{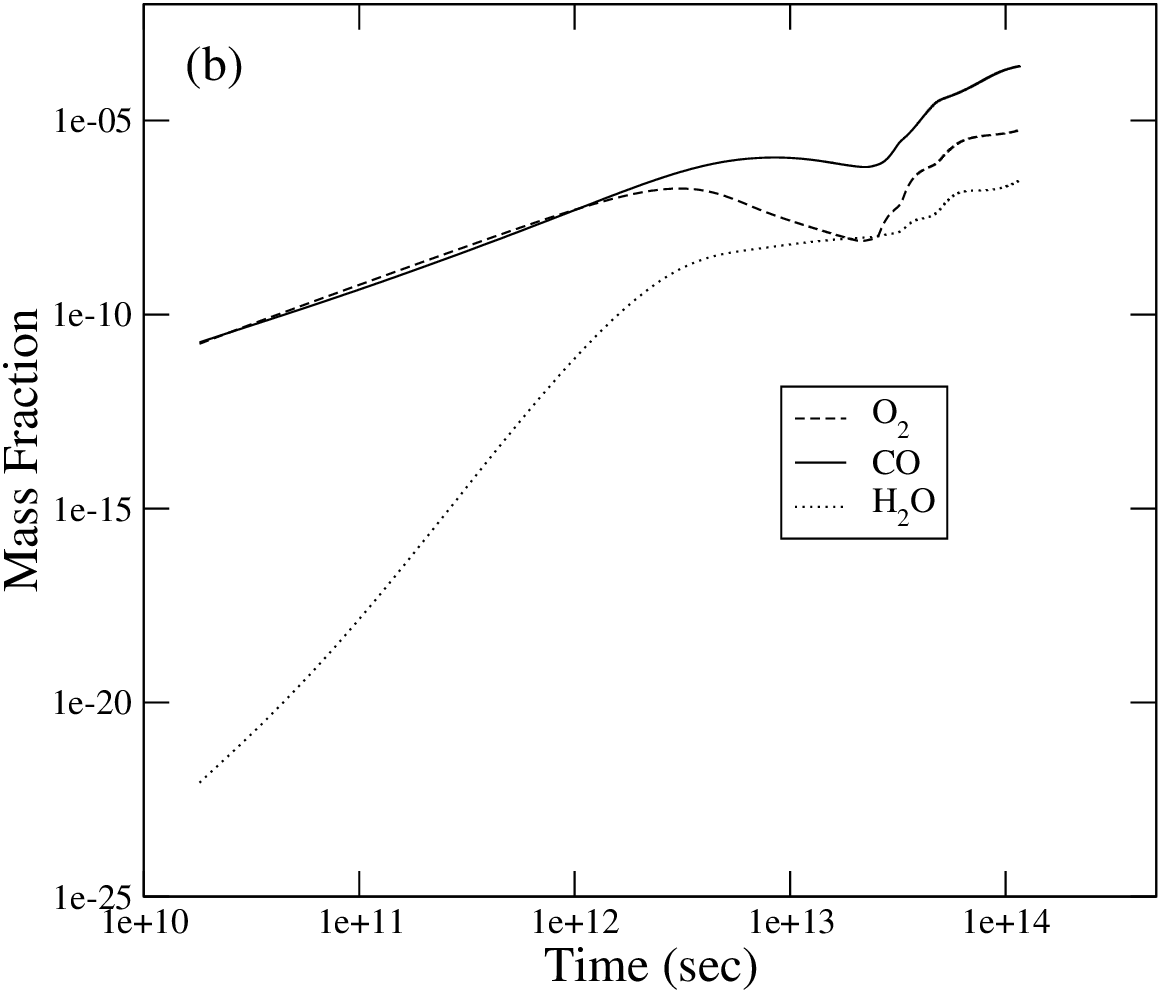} \\[1cm]
\includegraphics[width=.4\linewidth]{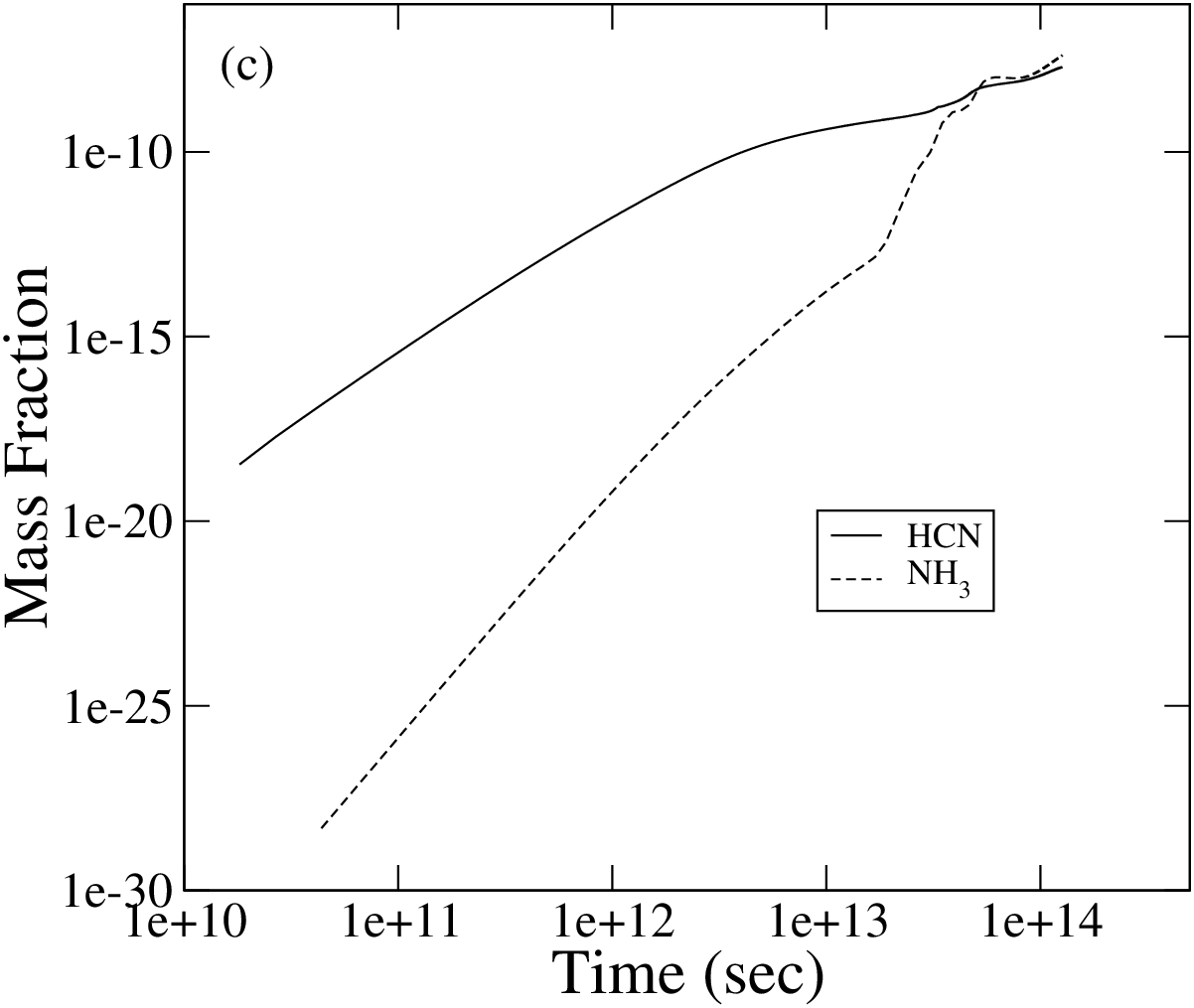} &&
\includegraphics[width=.4\linewidth]{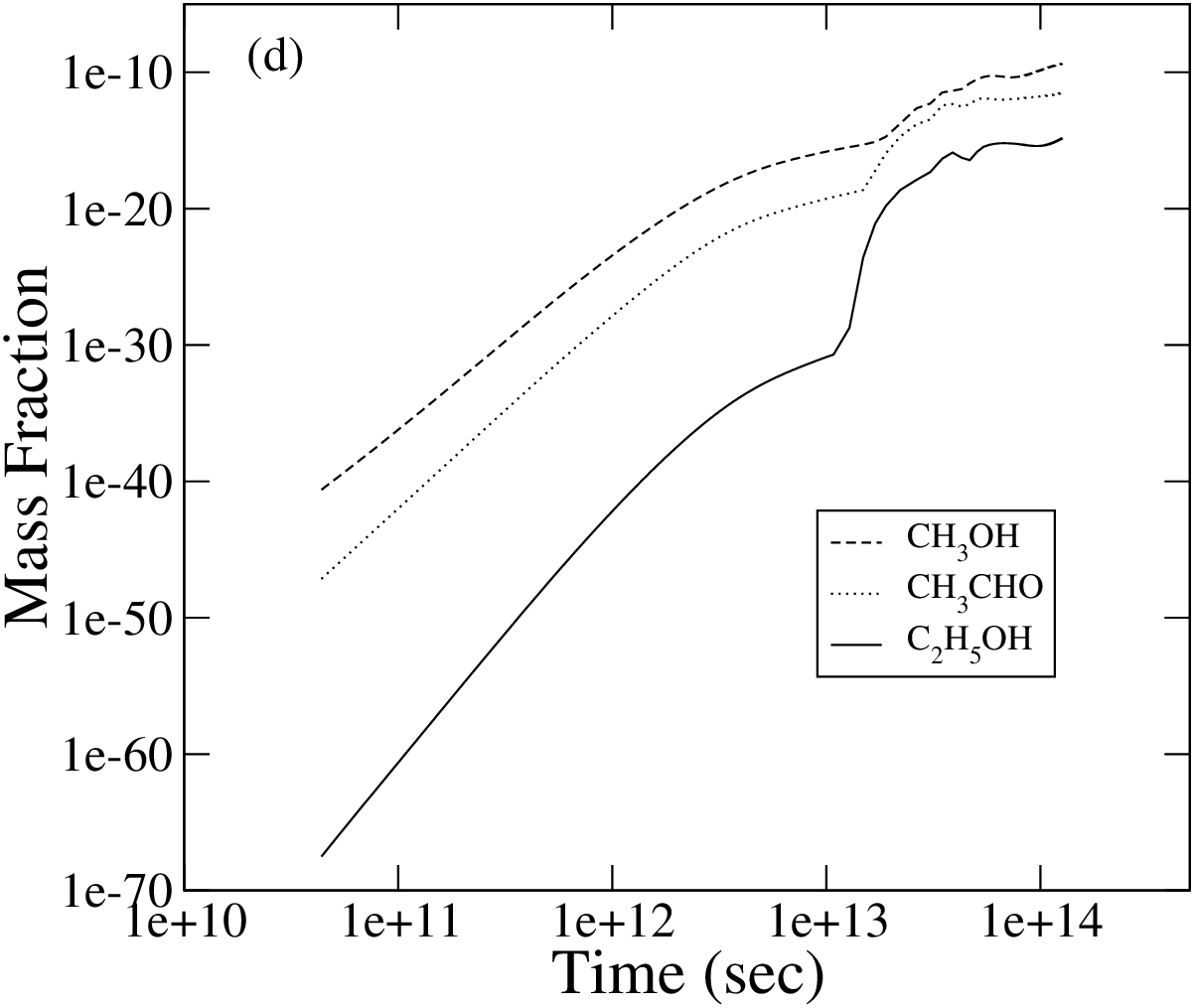}
\end{tabular}
\caption{Time variation of the spatially averaged mass fractions 
of (a) $H$ and $H_2$, (b) $O_2$, $H_2O$ and $CO$ (c) $HCN$ and $NH_3$ and (d) 
$C_2H_5OH$, $CH_3CHO$ and $CH_3OH$. Here, Model A cloud is chosen and the 
initial mass fractions are the same as that of Set 1. }
\end{figure}

We now show the variation of different species for Model A hydrodynamic
simulation. Figures 3(a-d) show the time variation of the average abundances 
of several species using the initial composition as in 
Miller et al. (1997) (Set 1). In Figure 3a we show the evolution of $H$ and $H_2$. 
The mass fraction of the atomic hydrogen goes down since it is utilized to form $H_2$ and other 
hydrogenated species. Since the atomic hydrogen is very much reactive agent it
goes to the molecular form easily and thus towards the end of our simulation almost all
the $H$ was found primarily in the form of $H_2$. In Figure 3b, we show the 
abundances of $O_2$, $H_2O$ and $CO$ whose final mean mass fractions are
around $10^{-5}$, $10^{-7}$ and $10^{-3}$ respectively. The production rate 
suddenly increases after the grids are filled up with the collapsing shell 
of matter and the densities go up. In Figure 3c, the time dependences of the 
abundances of $HCN$ and $NH_3$ are shown. The final mass fractions of these 
species are around $\sim 10^{-7}$. In Figure 3d, we show the variations of 
$C_2H_5OH$, $CH_3CHO$ and $CH_3OH$ whose final mass fractions
are $\sim 10^{-15}$, $\sim 10^{-10}$ and $\sim 10^{-9}$ respectively. 
Finally, in Figure 4, we show the evolutions of some of the simple bio-molecules,
such as glycine and alanine whose final mass fractions are found to be 
around $\sim 10^{-13}$ to $\sim 10^{-15}$ respectively.

\subsection{Collapse of clouds with other initial 
carbon abundances and comparison with observations}

\begin {figure}
\centering
\includegraphics[width=.4\linewidth]{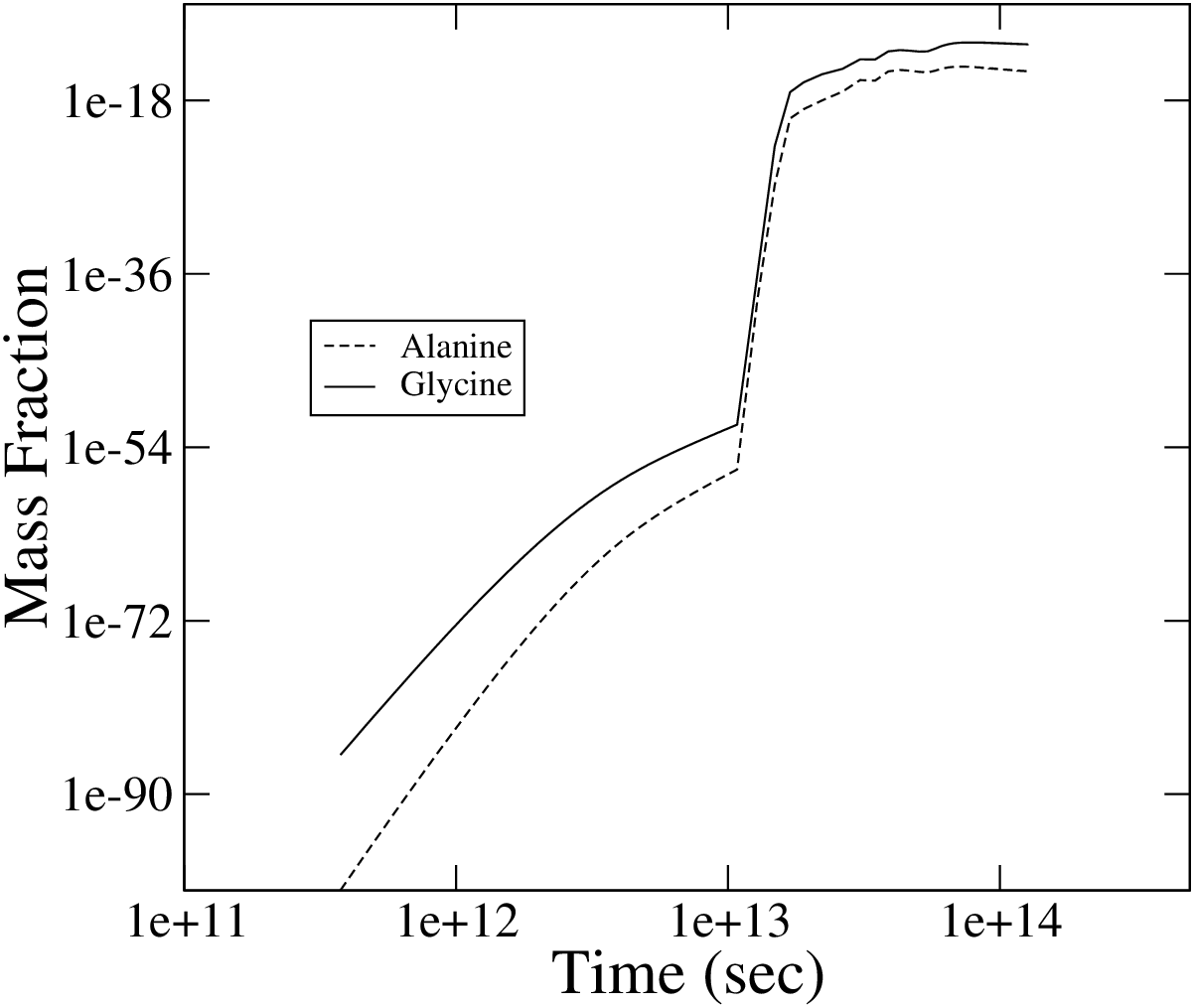}
\caption{Time variation of the spatially averaged mass fractions of two
simple bio-molecules, such as glycine and alanine. The hydrodynamic and chemical
model are the same as in the previous Figure.}
\end{figure}

\begin {figure}
\centering
\includegraphics[width=.4\linewidth]{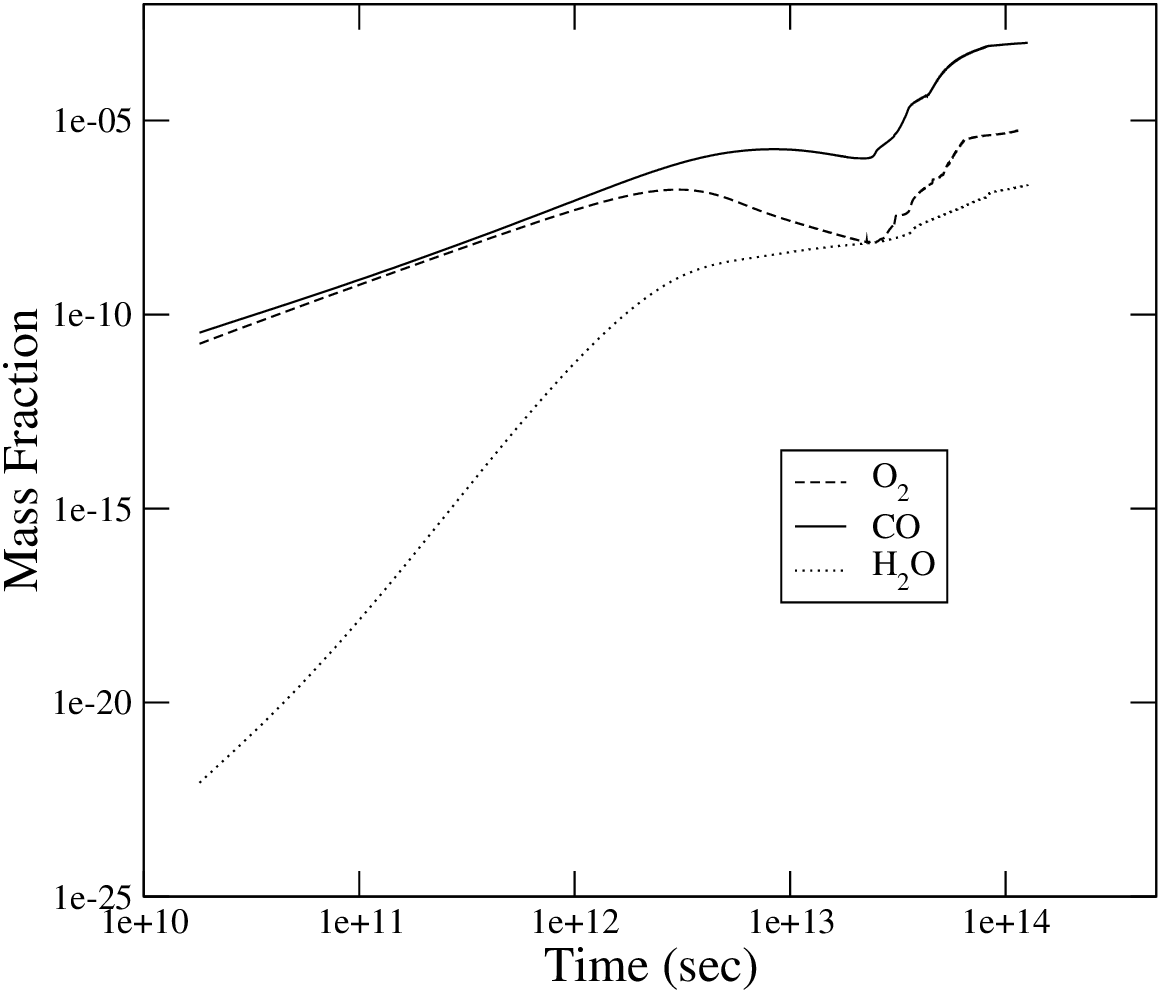}
\caption{Time variation of the spatially averaged mass fractions of
$H_2O$ and $CO$ when the initial carbon abundance is twice that in Set 1. }
\end{figure}

Since the initial composition of the clouds is not known very accurately, 
we ran Model A for one more initial chemical abundance (Set 2), where we 
use the mass fraction of carbon to be $0.001$, roughly twice as much as we 
used in Set 1 above. This may mimic the environment of an evolved star. 
We adjusted the mass fraction of helium accordingly so as to keep the sum 
of the mass fractions to be unity. As expected, the carbon
containing species become more abundant in simulations with  Set 2 as the
initial abundance. In Figure 5, we present the variations of $O_2$, $H_2O$ 
and $CO$ with time for these two models. These are to be compared with those 
presented in Figure 3b. 

\begin {figure}
\centering
\begin{tabular}{ccc}
\includegraphics[width=.4\linewidth]{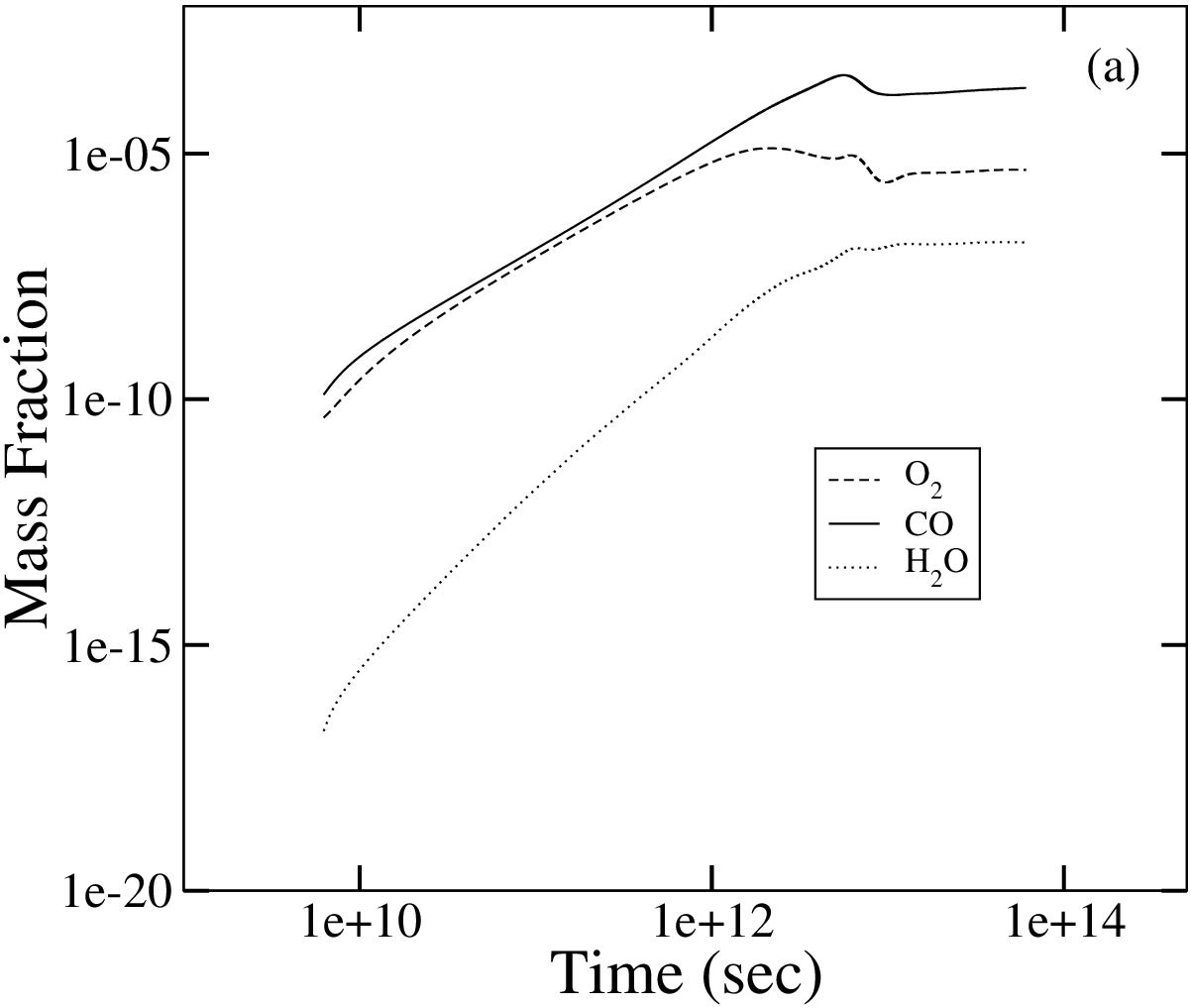} & \hspace{1cm} &
\includegraphics[width=.4\linewidth]{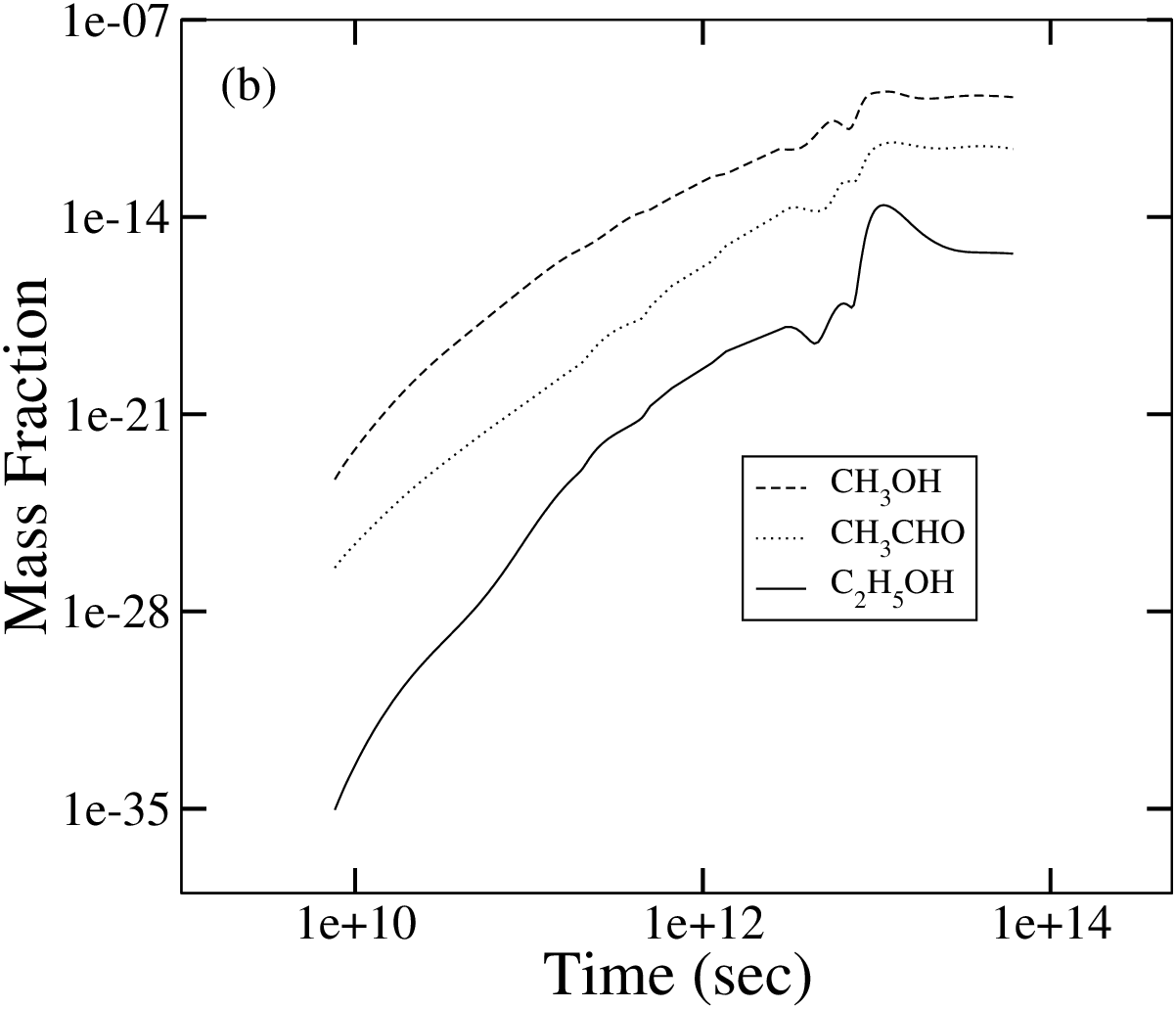}
\end{tabular}
\caption{Time variation of the spatially 
averaged mass fractions
of (a)$H_2O$ and $CO$ (b)$C_2H_5OH$, $CH_3CHO$ and $CH_3OH$. Here Model B1
was chosen. }
\end{figure}

\begin {figure}
\centering
\begin{tabular}{ccc}
\includegraphics[width=.4\linewidth]{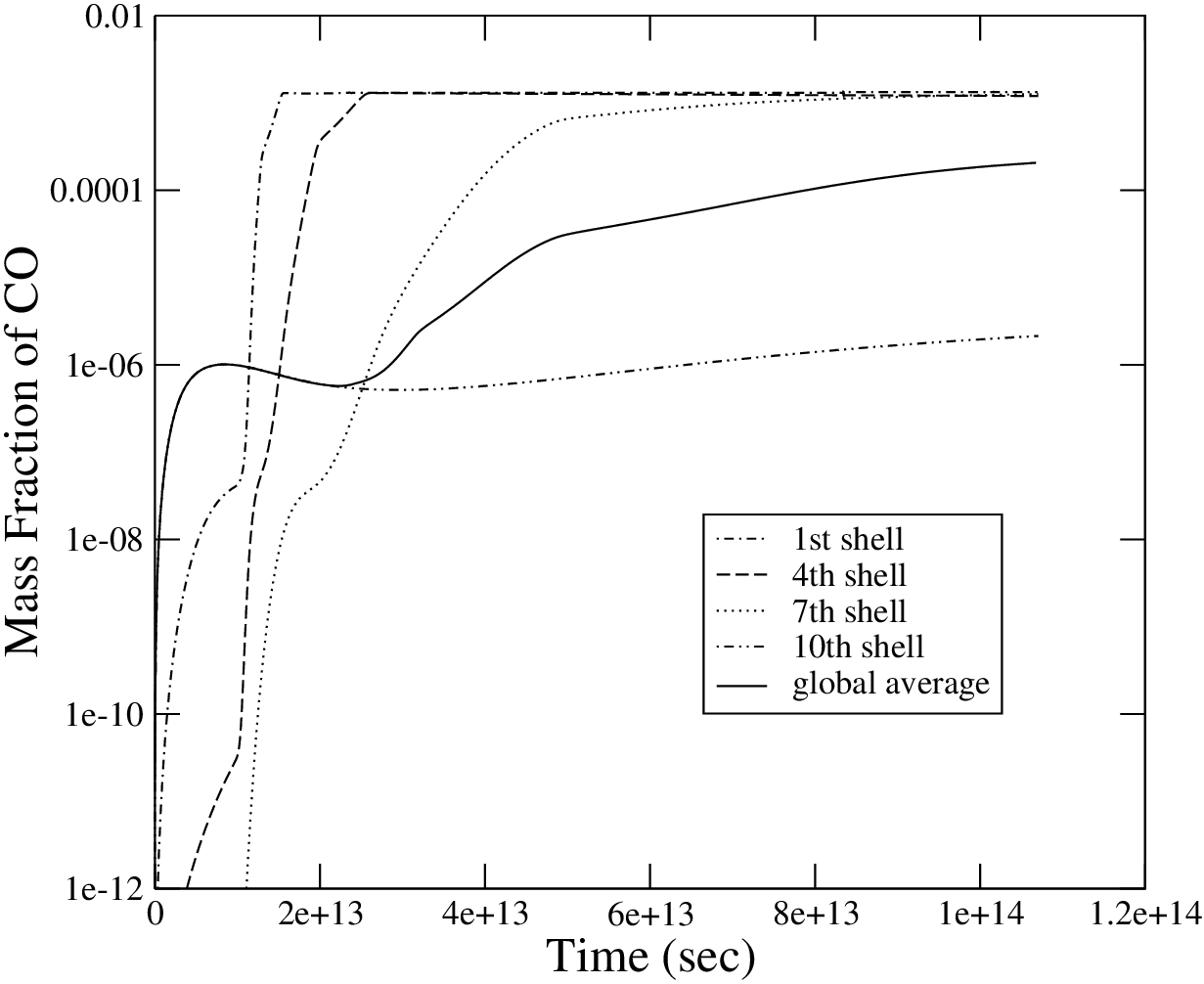} & \hspace{1cm} &
\includegraphics[width=.4\linewidth]{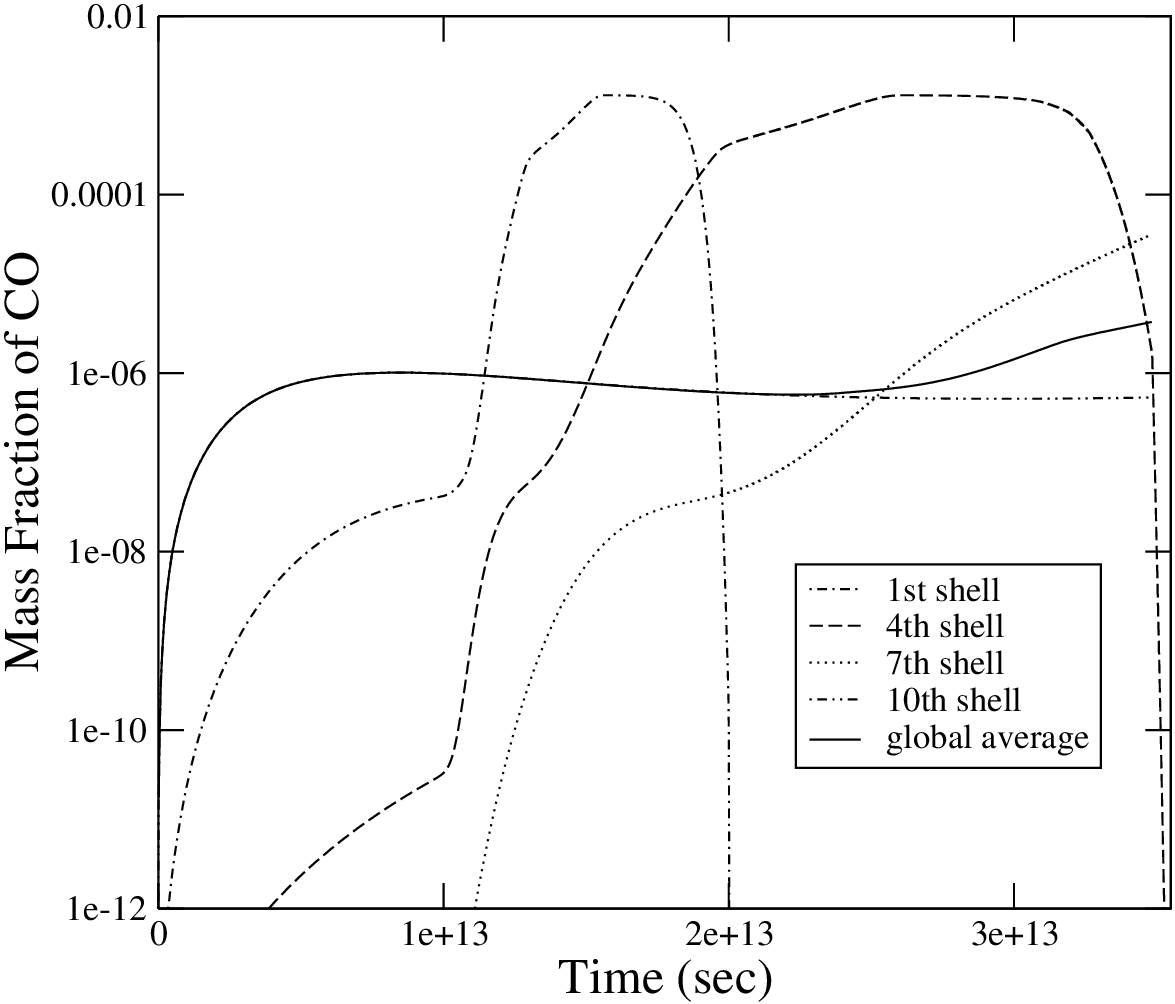}
\end{tabular}
\caption{Evolution of the mass fraction of CO (a) without 
and (b) with the freeze-out effect taken into account. The results in the 1st, 4th, 7th 
and 10th shells are displayed. Note that in the 1st and the 4th shell, the 
freeze-out effect is prominent. However, the global averages in both the 
cases do not differ by any large margin since they are dominated by the outer 
shells. Evolution in (b) only up to the freeze-out time is depicted to show 
detailed variation.}
\end{figure}

Figures 6(a-b) show the time variation of the average abundances of several
species using the initial composition of Model B1. 
In this connection, we wish to note that though we have been using the 
so-called freeze-out of the chemical species, the effect is not very easy to 
see in our abundance plots. This is because we are plotting the globally 
averaged abundances which is dominated by very slowly evolving
matter in the outermost cell. By the time the matter is dense and more 
evolved, it is compressed to a smaller volume and does not contribute much 
in the weighted average. However, when we plot the evolution for each cell, 
we see that the freeze-out effect can be seen
in the innermost four cells. In Figures 7(a-b) we show the evolutions of 
the mass fraction of CO in the 1st (innermost), 4th, 7th and 10th (outermost) 
shells of matter in the cloud when (a) we do not consider the freeze-out 
effect and when (b) we do consider the freeze-out effect.
When no freeze-out is considered (Figure 7a), all the shells eventually reach 
a saturation value of the mass-fraction. This is reached when the enhancement 
of the CO through combination of C and O matches its reduction 
through hydrogenation. Different shells take different times depending on the
average density of each shell. When we assume a freeze-out, the mass 
fractions in the innermost four shells start to decrease after reaching the 
saturation while the other shells do not decrease within the evolution time 
of the cloud. The evolution of the global average follows initially the 
evolution of the 10th shell, but afterwards, from around $t=2\times 10^{13}$s, 
the deviation occurs as the global average is dominated by denser 
intermediate shells. 

In Table 1, we present the results of Model A with Sets 1 and 2 chemical 
compositions, and the result from Model B using Set 1 composition (i.e., Model B1). 
We also compare the observed abundances of some of the species. For 
comparison, we converted mass fraction of a species into ratio of the number 
density of that species and that of $H_2$ so that our results are directly 
comparable with observables. The references from where the
observed abundances have been obtained are provided in Column 6.

\begin{table*}
\scriptsize
  \caption{Comparison of our Results with the observed abundances}
  \begin{tabular}{@{}llrrrrlrlrlr@{}}
  \hline
  \hline
 Molecules   & Observed   &\multicolumn{2}{c}{Model A}& Model B& &
References\\
\hline
\hline
$^aHCN$ &$ 6 \times 10^{-9}$&$1.4 \times 10^{-9}$&$2.4 \times 10^{-9}$ &
$1.9 \times 10^{-9}$&& Irvine and Hjalmarson (1983)  \\
$^aNH_3$&$ 5 \times 10^{-8}$&$4.5 \times 10^{-9}$&$1.2  \times 10^{-9}$&
$3.9 \times 10^{-9}$&&  T\"olle et al. (1981) \\
$^aCO$ & $ 3 \times 10^{-5}$ & $1.8 \times 10^{-5}$&$6.6  \times 10^{-5}$&
$2.6 \times 10^{-5}$&& Allen and Knapp (1978)\\
$^aH_2O$ & $7\times 10^{-8}$  & $4 \times 10^{-8}$&$2.2 \times 10^{-8}$&
$1.6 \times 10^{-8}$&& Snell et al. (2000) \\
$^aO_2$ & $10^{-6}$  & $3.3 \times 10^{-7}$&$1.9 \times 10^{-7}$&
$2.7 \times 10^{-7}$&& Gold et al. (2000) \\
$^aCH_3OH$& $5 \times 10^{-10}$ &$2.5 \times 10^{-11}$&$4.4\times 10^{-11}$ &
$1.6 \times 10^{-11}$&& Friberg et al. (1988)\\
$^aCH_3CHO$&$3 \times 10^{-10}$& $1.3 \times 10^{-13}$&$9.1 \times 10^{-13}$&
$1.9 \times 10^{-13}$&& Matthews et al. (1985)\\
$^{b}C_2H_5OH$&$1.5\times 10^{-9}$&$5.8 \times 10^{-17}$&$3.6 \times
10^{-16}$&$1.1 \times 10^{-15}$&& Ohishi et al. (1995) \\
Alanine& - & $2.3 \times 10^{-17}$&$8.3 \times 10^{-17}$&$6 \times 10^{-17}$&
& -
\\
Glycine& $10^{-10}$ $^c$  & $1.7 \times 10^{-14}$&$2.9 \times 10^{-14}$&
$1.7 \times 10^{-14}$&& Ceccarelli et al. (2000)\\
& $7\times 10^{-9}$ $^d$  &&&&& and\\
& $(0.21-1.5)\times 10^{-9}$ &&&&& Kuan et al. (2003)\\
                                                                                                              
\hline
\hline
\end{tabular}\\
$^a$ Observed near \astrobj{TMC-1} (10K)\\
$^{b}$ Observed near \astrobj{Orion KL} (70K)\\
$^c$ upper limit (cold cloud) ; $^d$ upper limit (hot core)\\
\end{table*}

From the Table, it is clear that our results in Sets 1-2 are
generally close to the observational results for most of the lighter species.
Since all the C and O are primarily channeled into the formation of CO,
production of other and more complex hydrocarbons by our method is not
very efficient. For instance, for $C_2H_5OH$, glycine, alanine, 
abundances obtained by our simulations are very low compared to the 
observed or 'upper limits' obtained from observations. According to Ohishi et al. (1995),
$C_2H_5OH$ is difficult to form in the gas phase. 
This is thought to be synthesized on the grain surfaces and then evaporated. 
Thus we believe that there could be other pathways than the method adopted by 
us to form glycine. These may be some of the reasons of the discrepancy 
seen in this Table.

\section{Concluding remarks}

In this paper, we presented the preliminary results on the
chemical evolution inside a collapsing interstellar cloud.
Our models are distinctly different from the other models used by 
several authors (Aikawa et al., 2005; Ceccarelli et al., 1996; Lim et al., 1999;
Shalabiea and Greenberg, 1995; Shematovich et al., 1997).
We do not include the heating and cooling processes while determining 
the dynamics of the cloud, and thus our model is not totally self-consistent. 
While in the outer edge of a diffused cloud 
the heating and cooling time scales are comparable to the infall time scales
and should have been included, deep inside, the infall time scale is much shorter
and cooling can be ignored. On the other hand, the 
outermost shell is also of very low density
and thus the heating is low. The cooling is also negligible as the reactions 
rates are low. Thus we believe that even if the heating and cooling 
were included, the result would not have differed significantly.

Unlike the previous study 
Chakrabarti and Chakrabarti (2000a, 2000b) we have incorporated the grain
chemistry of $H_2$ formation self-consistently. This major
improvement gave the most realistic abundances of $H_2$ molecules in
the grain and the gas phases. We find that our computed average abundances
generally agree with the observed abundances (see, e.g., Allen and Knapp, 1978;
Friberg et al., 1988; Irvine and Hjalmarson, 1983; Matthews et al., 1985; 
Ohishi et al., 1995; T\"olle et al., 1981)
except when the molecules are complexes. We
always seem to underestimate  them as compared to the observed values or 
`upper limits'. It is not clear at this moment how close the results of 
the bio-molecules are in comparison to the actual values, since the 
reaction rates we used are similar to the neutral-neutral rates which 
need not be accurate. If, for instance, the reaction rate at each 
step of glycine formation were higher by a factor of ten,
the resulting glycine abundance would have been $10^3$ times higher, which is
closer to observed claims. Given that the pathways to produce glycine may 
itself be different, perhaps one needs to look into laboratory experiments 
which simulate interstellar clouds for guidance (see, e.g., Elsila, 2007).
                                                              
In future, we plan to improve the hydrodynamic model by including the angular
motion and shock formation in the flow. We expect that jets and outflows
would form and a part of this will fall back on the disk and the matter
would be recycled. We also plan to improve the grain chemistry to include
the formation of $CH_3OH$, $CO$, $NH_3$, $OH$ on the grains themselves.
Thus we anticipate that the chemical abundance will be strongly affected
by such recycling of matter and such incorporation of newer species on the
grain surfaces.

A. Das acknowledge the support from an ISRO project.

\newpage
\vskip 2cm
\hskip 6cm {\Large \bf APPENDIX}
\vskip 0.5cm
\hskip 5cm {\bf SOLUTION PROCEDURE}\\
The equations are solved on a spherical grid extending from $r_{in}$ to 
$r_{out}$ composed of $N$ equal logarithmically spaced grids along the radial 
direction. The code is customized to take care of the collapsing spherical 
hydrodynamical flow which is strictly one dimensional. That is, no back flow 
is possible. Thus, to solve Eqs. (1-2) we use the first order upwind 
differencing method. After appropriately splitting, the Eqs. (1-2) become,
$$
\rho _{i}^{j+1}=\rho_{i}^{j} - \frac{dt}{{r_i}^2(r_{i+1}-r_i)}
(\rho _{i+1}^{j}v{_r}_{i+1}^{j}
r_{i+1}^2-\rho _{i}^{j}v{_r}_{i}^{j}r_{i}^2),
\eqno (A.1)
$$
$$
\rho_{i}^{j+1}v{_r}_{i}^{j+1}= \rho_{i}^{j}v{_r}_{i}^{j} -
\frac{dt}{r_{i}^2 (r_{i+1}-r_i)}
$$
$$
({\rho _{i+1}^{j}}^2v{{_r}_{i+1}^{j}}^2r_{i+1}^2/\rho_{i+1}^{j}-
{\rho _{i}^{j}}^2{v{_r}_{i}^{j}}^2r_{i}^2/\rho_{i}^{j})
$$
$$
-\frac{dt}{(r_{i+1}-r_i)} [\rho_{i}^{j}(\phi_{i+1}^{j}-\phi_{i}^{j})+
(p_{i+1}^{j}-p_{i}^{j})]
\eqno (A.2)
$$
Here, $i$ denotes the index for the radial grid and $j$ denotes the index 
for the time. To avoid the instability in the code we chose the time step by
using the Courant-Friedrichs-Lewy stability criterion, which gives,
$$
{|v|\Delta t}/{\Delta r} \leq 1,
\eqno (A.3)
$$
i.e.,
$$
\Delta t \sim \Delta r/|v|,
$$
where, $|v|$ is the magnitude of velocity, $\Delta t$ is the time step,
and $\Delta r$ is the grid spacing along the radial direction.
We always advance the time step after ensuring that the Courant condition
is satisfied. To be on the safer side, we chose time step $dt=\Delta t/2$.
Even though we use self-gravitating flow, we do not solve Poisson equation 
to get the potential $\phi_i (r)$ here, since we are dealing with a spherical 
flow. Potential at any point is computed as a sum of two terms, one coming 
from the cloud itself [$\phi_{cloud}= -GM_{cloud}(r)/r$, where, 
$M_{cloud}(r)$ is the mass of the cloud $r_{in}<r<r_{out}$ within the grid]
and the other is due to the cloud core $\phi_{core}=-GM_{core}/r$,
where, $M_{core}$ is the mass of the core within $r<r_{in}$, the inner edge 
of the computational grid. The value of $\phi (r)$ that is required during 
the simulation is dynamically calculated
from the mass within $r$, i.e., $\phi=-G(M_{cloud}+M_{core})/r$. 
Here, $M_{cloud}$ at $j$th time step is calculated by adding contributions 
from each spherical shell, 
$$
{M_{cloud}}^j=\Sigma_i 4\pi {r_i}^2 dr_i \rho_{i}^{j},
\eqno (A.4)
$$
where, $i=1,2,..... N$ and the densities at the $j$th time step are used. 
We put $r=r_{out}$ to get the potential at the outer  boundary.

{}
\end{document}